\documentclass{revtex4}

\usepackage{amsmath,amsthm}
\usepackage{rotating}
\renewcommand{\arraystretch}{2}

\usepackage{graphicx}

\usepackage{amsfonts}
\newtheorem{theorem}{Theorem}[section]
\newtheorem{lemma}[theorem]{Lemma}
\newtheorem{proposition}[theorem]{Proposition}
\newtheorem{cor}[theorem]{Corollary}

\theoremstyle{remark}
\newtheorem{remark}[theorem]{Remark}

\theoremstyle{definition}
\newtheorem{definition}[theorem]{Definition}

\theoremstyle{example}
\newtheorem{example}[theorem]{Example}

\theoremstyle{notation}

\newcommand{\bra}[1]{\langle#1|}
\newcommand{\ket}[1]{|#1\rangle}

\begin{document}

\title{Effective approach to open systems with probability currents and the Grothendieck formalism}           
\author{A. Vourdas}
\affiliation{Department of Computer Science,\\
University of Bradford, \\
Bradford BD7 1DP, United Kingdom\\a.vourdas@bradford.ac.uk}

\begin{abstract}
An effective approach to open systems and irreversible phenomena is presented, where an open system $\Sigma(d)$ with $d$-dimensional Hilbert space, is a subsystem of a larger isolated system $\Sigma(2d)$ 
(the `full universe') with $2d$-dimensional Hilbert space.
A family of Bargmann-like representations  (called $z$-Bargmann representations) introduces naturally the  larger  space. 
The $z$-Bargmann representations are defined through semi-unitary matrices (which are a coherent states formalism in disguise).
The `openness' of the system is quantified with the probability current that flows from the system to the external world.
The Grothendieck quantity ${\cal Q}$ is shown to be related to the probability current, and is used as a figure of merit for the `openness' of a system. 
${\cal Q}$ is expressed in terms of `rescaling transformations' which change not only the phase but also  the absolute value of the wavefunction, 
and are intimately linked to irreversible phenomena (e.g., damping/amplification).
It is shown that unitary transformations in the isolated system $\Sigma(2d)$ (full universe), reduce to rescaling transformations when projected to its open subsystem $\Sigma(d)$.
The values of the Grothendieck ${\cal Q}$ for various quantum states in an open system, are compared with those for their counterpart states in an isolated system.

\end{abstract}
\maketitle

\section{Introduction}

Open systems and irreversible phenomena have been studied for a long time (e.g., \cite{OP0,OP1,OP2,OP3,OP4,D1,D2}). 
They use a tensor product of the Hilbert space of the open system with the Hilbert space of the environment (which we call `tensor product environment').
Various approximations in this context have been studied extensively (quantum Markov processes, the Gorini-Kossakowski-Sudarshan-Lindbland equation\cite{GKS,L}, etc).
In this context the Kraus formalism \cite{KR} provides a nice mathematical language for open systems.
All these approaches use tensor products, and we call them `tensor product formalisms' .
Similar ideas are used currently in the developing area of quantum thermodynamics (e.g., \cite{T1,T2}). 

In this work we present a `direct sum formalism' for open systems which
introduces a probability current from the system to the external world, and also links this approach to the Grothendieck formalism.
In an open system we have a probability current coming out of the system, and also the absolute value of the 
wavefunction might increase (amplification) or decrease (damping) as a function of time. 
For this reason we regard  the open system as a subsystem of a larger isolated system (the `full universe'). 

Using a `Bargmann-like' representation, an open system  $\Sigma(d)$ with $d$-dimensional Hilbert space $H(d)$, becomes in a natural way a subsystem of a 
larger isolated system $\Sigma(2d)$ with $2d$-dimensional Hilbert space $H(2d)$.
The `Bargmann-like' representation represents the states in $H(d)$ with
$2d$-dimensional vectors in $H(2d)$. $H(d)$ is embedded in a natural way into $H(2d)$,
and then there is `external world' where the probability current coming out of $\Sigma(d)$ can go (Eq.(\ref{700})).

This is an effective formalism where the Hilbert space of the full universe is the direct sum of the Hilbert space of the system and the Hilbert space of an `external system',
which we call `direct sum environment' and denote as $\Sigma_{\rm ext}(d)$.  The `direct sum environment'  is a very different concept from the `tensor product environment'. 
We call the present approach direct sum formalism. It avoids the complexity of the tensor product formalisms, and at the same time allows for irreversible phenomena in $\Sigma(d)$.
We quantify how open the system is, with a probability current from the system to the external system.

The Bargmann representation is usually linked to coherent states. Here we introduce $2d$ coherent states in $H(d)$, and $d\times 2d$ `semi-unitary matrices' which have them as columns.
The properties of semi-unitary matrices are intimately related to the properties of coherent states (e.g., a resolution of the identity).
In this sense semi-unitary matrices are a coherent states formalism in disguise.

In this general formalism for open systems and irreversible phenomena, we use as a figure of merit the Grothendieck quantity ${\cal Q}$.
The Grothendieck inequality in pure mathematics, was originally formulated\cite{QR1} in the context of a tensor product of Banach spaces.
For this reason applications in quantum physics\cite{QR2,QR3,QR4,QR5,QR6,QR7,QR8,QR9,QR10}  have been in entangled multipartite systems described by tensor products of Hilbert spaces.
In this context there are deep links between the Grothendieck inequality and the violation of Bell-like inequalities. 
Later mathematical work \cite{GR1,GR2,GR3,GR4,GR5,GR6} emphasised that the Grothendieck inequality can be formulated without tensor products.
This motivated our work\cite{VOU1,VOU2} that uses the Grothendieck formalism in a single quantum system.

In a physical language the Grothendieck formalism starts with a `classical' quadratic form ${\cal C}$ that uses complex numbers (scalars) in the unit disc.
It then replaces the scalars with vectors in the unit ball of a Hilbert space (passage from classical to quantum mechanics) and considers  a `quantum' quadratic form ${\cal Q}$.
The Grothendieck inequality states that if the classical ${\cal C}\le 1$, the corresponding quantum ${\cal Q}$ might take values greater than $1$, up to the complex Grothendieck constant $k_G$.
We are especially interested in the  region  ${\cal Q}\in (1,k_G)$ because  the quantum ${\cal Q}$ can take values in it, while the classical ${\cal C}$ cannot  take values in it (it is classically forbidden region). 
An example of this is discussed in section \ref{RR2}.

The general area of irreversible phenomena and open systems needs novel quantities, which are generalisations of the ones used in isolated systems.
For example, isolated systems are described with unitary transformations, whilst irreversible phenomena need more general transformations that can change not only the phase but also the absolute value of 
the wavefunction (amplification/damping), and we recently studied them under the name rescaling transformations \cite{VOU4}. 
${\cal Q}$ is expressed in terms of rescaling transformations in Eq.(\ref{GR1}) and this shows its link to irreversible phenomena (e.g., damping/amplification).

The present paper uses a direct sum approach to open systems (as opposed to existing tensor product approaches to open systems). In particular:
\begin{itemize}
\item
We present the Grothendieck quantities  as traces of products of matrices that involve rescaling transformations.
We also show that in special cases ${\cal Q}$ reduces to many well known physical quantities (e.g., the Weyl and Wigner functions in Eqs.(\ref{W1}),(\ref{W2})).
\item
We introduce a family of  $d\times 2d$ semi-unitary matrices which are related to coherent states, and study their properties.
\item
We define $z$-Bargmann representations using semi-unitary matrices ($z$ is a parameter with $|z|=1$).
They enlarge the Hilbert space in a natural way (Eq.(\ref{780})), and introduce an `external world' which is needed for an open system.
 They are used to embed in a natural way, an open system $\Sigma(d)$ into a larger isolated system $\Sigma(2d)$ (`full universe').
  The constraints of an isolated system (conservation of probability, invariance of physical quantities under unitary transformations, etc) hold in 
$\Sigma (2d)$, but do not hold in its open subsystem $\Sigma (d)$. 
 \item
 The `openness' of the system is quantified with the probability current from the open system to the external world.
 Numerical results show a quasi-periodic exchange between the system and the environment, with the probability current taking alternately positive and negative values. 
 The formalism has the parameter $z$, and different values of $z$ describe different coupling of the system to the external world.
 The probability current depends on $z$.
  \item 
 We show that  unitary transformations in the `full universe' (which is an isolated system) are projected into rescaling transformations in the open subsystem (section \ref{RRR}).
 \item
 We show that the time derivative of the Grothendieck ${\cal Q}$ is related to the probability current from the open system to the external world (section \ref{RR1}).
 \item
 We give an example where ${\cal Q}>1$ (section \ref{RR2}).
 We also compare the values of ${\cal Q}$ and $\frac{d{\cal Q}}{dt}$ for states in an open system, with  their counterparts in an isolated system, and show that in general they are different  (section \ref{sec67}).
 
\end{itemize}

 In section 2 we define the notation and present briefly a simplified view of some of the concepts, within a `toy model'.
 In section 3 we summarise briefly the Grothendieck formalism with emphasis on its relation to rescaling matrices and irreversible phenomena.
Section 4 introduces semi-unitary matrices, which are related to coherent states.

Section 5 uses the semi-unitary matrices to define a family of Bargmann-like representations which we call $z$-Bargmann representations ($z$ is a constant on the unit circle).
They are used to embed the Hilbert space $H(d)$ of $\Sigma(d)$, into the larger Hilbert space $H(2d)$ of $\Sigma(2d)$. 
 Section 6 discusses probability currents, and gives numerical results for various examples.
 In section 7 we present various results which show that the Grothendieck ${\cal Q}$ is a figure of merit for the `openness' of a system. 
 We conclude in section 8 with a discussion of our results.

\section{Preliminaries}
\subsection{Notation}

We consider a quantum system $\Sigma(d)$ with variables in ${\mathbb Z}_d$ (the ring of integers modulo $d$), and a $d$-dimensional Hilbert space $H(d)$.
$|X;\nu\rangle$ is an orthonormal basis which we call position basis (the $X$ in this notation is not a variable, but it simply indicates position basis).
\begin{eqnarray}\label{1}
\ket{X; \nu}=\begin{pmatrix}
0&\dots&1&\cdots &0
\end{pmatrix}^\dagger;\;\;\;0\le\nu\le d-1.
\end{eqnarray}
Here all rows are zeros except the $\nu+1$ row which is one.

Through a Fourier transform we get another orthonormal basis that we call momentum basis:
\begin{eqnarray}
\ket{P;\nu} =F\ket{X;\nu};\;\;\;\;F=\frac{1}{\sqrt d}\sum _{\mu,\nu}\omega^{\mu \nu}\ket{X;\mu}\bra{X;\nu};\;\;\;\omega=\exp \left (i\frac{2\pi }{d}\right ).
\end{eqnarray}

Let $X$ be the $d\times d$ `upwards displacement' matrix (circulant permutation matrix):
\begin{eqnarray}\label{1X}
X=\begin{pmatrix}
0&1&0&\cdots&0\\
0&0&1&\cdots&0\\
\vdots&\vdots&\vdots&\ddots&\vdots\\
0&0&0&\cdots&1\\
1&0&0&\cdots&0
\end{pmatrix};\;\;\;X^\mu\ket{X;\nu}=\ket{X;\nu-\mu};\;\;\;X^d={\bf 1}_d;\;\;\;{\bf 1}+X+X^2+...+X^{d-1}=J_d
\end{eqnarray}
Here $J_d$ is the `matrix of ones' (all elements are equal to $1$). 
${\cal G}_d=\{{\bf1}_d,X,...,X^{d-1}\}$ is a multiplicative group isomorphic to the additive group ${\mathbb Z}_d$.
We also consider the $d\times d$ matrix
\begin{eqnarray}\label{ZZ}
&&Z={\rm diag}
\begin{pmatrix}
1&\omega&\cdots&\omega^{d-1}
\end{pmatrix}
\end{eqnarray}
Then
\begin{eqnarray}\label{HW}
&&X^d=Z^d={\bf 1}_d;\;\;\;ZX\omega=XZ
\end{eqnarray}
The matrices 
\begin{eqnarray}\label{D}
{\mathfrak D}(a, b, c)=Z^aX^b\omega^c;\;\;\;a,b,c\in{\mathbb Z}_d
\end{eqnarray}
are displacement operators in the ${\mathbb Z}_d\times {\mathbb Z}_d$ phase space, and form a representation of the Heisenberg-Weyl group.

We will use the following notation for the matrix $1$-norm of a matrix $\theta$
\begin{eqnarray}
||\theta_{rs}||_1=\sum_{r,s}|\theta _{rs}|.
\end{eqnarray}
In table \ref{t0} we summarise the main symbols in the paper.

\subsection{A `toy model' for our approach to open systems}

In this section, we introduce briefly within a `toy model',  three concepts which are important in this paper:
\begin{itemize}
\item
Rescaling transformations which are a generalisation of unitary transformations.
\item
Probability currents.
\item
A matrix normalisation which is related to ${\cal C}(\theta)$ and $g(\theta)$ in the Grothendieck formalism below.
\end{itemize}

We consider a quantum system with $n$-dimensional Hilbert $H(n)$.
We assume that can only `see' (do measurements, etc) a $d$-dimensional subspace $H(d)$ of $H(n)$, and that the rest of the Hilbert space is `invisible'.
 Let ${\cal P}$ be the projector to $H(d)$.
 Also let $\rho(t)=U(t)\rho_0[U(t)]^\dagger$ be a density matrix describing a quantum state in $H(n)$, which evolves in time with the unitary time evolution operator $U(t)$.
We can only `see' and measure the 
\begin{eqnarray}
R(t)={\cal P}\rho(t){\cal P}=V(t)\rho_0[V(t)]^\dagger;\;\;\;V(t)={\cal P}U(t).
\end{eqnarray}
Then:
\begin{itemize}
\item
The transformations $V(t)$ are not unitary, but they obey the property
\begin{eqnarray}
\sum_j|V_{ij}|^2=(VV^\dagger)_{ii}={\cal P}_{ii} \le 1.
\end{eqnarray}
We used here the fact that the diagonal elements of projectors are less or equal to one.
This relation is generalised into a definition of rescaling matrices in Eq.(\ref{4}).
\item
${\rm Tr}[\rho(t)]=1$ which is related to conservation of probability in $H(n)$.
But ${\rm Tr}[R(t)]$ is a function of time, and we call probability current $J(t)$ its time derivative:
\begin{eqnarray}
J(t)=\frac{d}{dt}{\rm Tr}[R(t)]=\frac{d}{dt}{\rm Tr}[\rho(t){\cal P}]=-\frac{d}{dt}{\rm Tr}[\rho(t)({\bf 1}_n-{\cal P})].
\end{eqnarray}
This flows between the subspace $H(d)$ and the rest of the space.
\item
$R(t)$ is Hermitian positive semidefinite matrix, and is only a part of the full density matrix $\rho(t)$. 
Some kind of normalisation of $R(t)$ is needed in calculations with it.
This can be done in various ways, but we should not use $\frac{R(t)}{{\rm Tr}[R(t)]}$  because that would give probability current zero, when it is physically clear that there is 
flow of probability between $H(d)$ and the rest of the space.

The Grothendieck approach is to calculate the supremum $g(R)$ of all quadratic forms $|\sum R_{ij}a_ib_j|$ for all  $|a_r|\le 1$ and  $|b_r|\le 1$, and use the$\frac{R}{g(R)}$.
Unitary transformations do not preserve the  $|a_r|\le 1$ and  $|b_r|\le 1$, and we compare it below with another supremum $g^\prime(R)$ 
of all quadratic forms $|\sum R_{ij}a_ib_j|$ for all  $\sum |a_r|^2\le d$ and  $\sum |b_r|^2\le d$.
The difference between the two, turns out to be important.
\end{itemize}

\section{The Grothendieck inequality in terms of rescaling and dequantisation matrices}

The Grothendieck inequality considers a $d\times d$ complex matrix $\theta$ and the `classical quadratic form'
\begin{eqnarray}\label{89}
{\cal C}(\theta)=\left |\sum_{r,s}\theta _{rs}a_rb_s\right |;\;\;\;|a_r|\le 1;\;\;\;|b_s|\le 1.
\end{eqnarray}
${\cal C}(\theta)$ is classical in the sense that the $a_r, b_s$ are scalars in the unit disc.

It also considers the corresponding `quantum quadratic form' where the scalars are replaced with vectors:
\begin{eqnarray}\label{34}
{\cal Q}(\theta)=\left |\sum_{r,s}\theta_{rs}\lambda_r\mu_s\bra{u_r}v_s\rangle \right |;\;\;\;\langle u_r\ket{u_r}=\langle v_r \ket{v_r}=1;\;\;\;\lambda_r, \mu_r\le1.
\end{eqnarray}
The $\lambda_r\ket{u_r}$, $\mu_s\ket{v_s}$ are vectors in the unit ball in $H(d)$.
${\cal Q}(\theta)$ is a `quantum quadratic form' in the sense that the scalars have been replaced with vectors.

The Grothendieck inequality states that if the classical ${\cal C}(\theta)\le 1$ (for all $a_r,b_s$ in the unit disc), then the quantum ${\cal Q}(\theta)$ 
might takes values greater than one, up to the complex Grothendieck constant  $k_G$:
\begin{eqnarray}\label{GR}
{\cal C}(\theta)\le 1\;(\forall\; |a_r|\le1;\;|b_s|\le 1)\implies{\cal Q}(\theta)\le k_G.
\end{eqnarray}
The exact value of $k_G$ is not known, but it is known that $k_G\le 1.4049$.

We note that ${\cal C}(\theta)$ in Eq.(\ref{89}) is not invariant of under unitary transformations. If $U$ is a unitary matrix, then the transformations
\begin{eqnarray}
\theta\rightarrow U\theta U^\dagger;\;\;\; b_s\rightarrow \sum _t U_{st}b_t;\;\;\;a_r\rightarrow \sum _t a_tU_{rt}^*,
\end{eqnarray}
will violate (in general) the requirement that $a_r, b_s$ should be in the unit disc.
In a quantum context unitary transformations are fundamental for isolated systems, 
and this is an early indication that the formalism is related to open systems.

Values of ${\cal Q}(\theta)$ in the region $(1,k_G)$ are important, because this is a classically forbidden region (the ${\cal C}(\theta)$ cannot take values in it).
In section \ref{RR2} we give an example with ${\cal Q}>1$.

We reformulated \cite{VOU1,VOU2} the Grothendieck inequality in terms of the trace of a product of matrices (which is more appropriate for quantum mechanics), as follows:
\begin{definition}
\mbox{}
\begin{itemize}
\item[(1)]

Let $V$ be a $d\times d$ matrix, and 
\begin{eqnarray}\label{4}
{\cal N}(V)=\max_i\sqrt{\sum_j|V_{ij}|^2}=\max_i\sqrt{(VV^\dagger)_{ii}}.
\end{eqnarray}
The set ${\cal S}_d$ of rescaling matrices, consists of matrices with ${\cal N}(V)\le 1$.
Any matrix $V$ after appropriate normalisation, belongs in ${\cal S}_d$: 
\begin{eqnarray}\label{434}
\lambda \frac{V}{{\cal N}(V)}\in {\cal S}_d;\;\;\;\lambda\le 1.
\end{eqnarray}
All unitary matrices are rescaling matrices  in ${\cal S}_d$.

\item[(2)]
A special case of rescaling matrices are the dequantisation matrices
\begin{eqnarray}
{\cal A}_{rs}(a_i)=\frac{1}{\sqrt{d}}a_r;\;\;\;|a_r|\le 1.
\end{eqnarray}
Here all elements in the same row are equal to each other. 
We call ${\cal T}_d$ the set of these matrices (${\cal T}_d\subset {\cal S}_d$).
\end{itemize}
\end{definition}
Rescaling  matrices acting on a vector multiply its length by a factor 
which can be contraction (describing phenomena like damping and quantum tunneling) or dilation (describing  amplification).
They are generalisations of unitary matrices which change only the phase of a vector and keep its length constant.

Dequantisation matrices map all vectors in the Hilbert space $H(d)$ into vectors in one-dimensional vector space (which can be viewed as scalars).
In this sense they convert a Hilbert space formalism into a formalism of scalars.

${\cal C}(\theta)$ in Eq.(\ref{89}) can be expressed 
in terms of dequantisation matrices ${\cal A}\in{\cal T}_d$ as
\begin{eqnarray}\label{23}
{\cal C}(\theta)=\left |{\rm Tr}\left [[{\cal A}(a_i^*)]^\dagger \theta {\cal A}(b_i)\right ]\right |.
\end{eqnarray}
Also
${\cal Q}(\theta)$ in Eq.(\ref{34}) can be written in terms of rescaling matrices $V,W\in {\cal S}_d$ as
\begin{eqnarray}\label{654}
{\cal Q}(\theta)=|{\rm Tr}(W^\dagger\theta V)|.
\end{eqnarray}
Here $V$ is a matrix that has the components of $\mu_s\ket{v_s}$ in the $s$-row and then $V\in {\cal S}_d$. Also $W$ is a matrix that has the 
components of $\lambda_r\ket{u_r}$ in the $r$-row and then $W\in {\cal S}_d$ (but $W^\dagger $ might not belong in ${\cal S}_d$).
Unitary matrices $V,W$ correspond to $\lambda_r=\mu_s=1$. We can easily show that:
\begin{eqnarray}
\lambda_r\mu_s\bra{u_r}v_s\rangle=(VW^\dagger)_{sr}.
\end{eqnarray}
For simplicity we use the notation ${\cal Q}(\theta)$, rather than the more accurate notation $Q(\theta, V, W)$ (the $V,W$ will be defined from the context).

The Grothendieck inequality in Eq.(\ref{GR}) becomes now
\begin{eqnarray}\label{GR1}
&&{\cal C}(\theta)=\left |{\rm Tr}\left [[{\cal A}(a_i^*)]^\dagger \theta {\cal A}(b_i)\right ]\right |\le 1\;(\forall \; |a_i|\le1;\;|b_i|\le 1)\implies {\cal Q}(\theta)=|{\rm Tr}(W^\dagger\theta V)|\le k_G\nonumber\\
&&{\cal A}(a_i^*), {\cal A}(b_i)\in {\cal T}_d;\;\;\;V,W\in {\cal S}_d.
\end{eqnarray}

\begin{definition}\label{def1}

For an arbitrary  $d\times d$ matrix $\theta$:
\begin{itemize}
\item[(1)]
$g(\theta)$ is the supremum of the set of the values of ${\cal C}(\theta)$ in Eq.(\ref{89}) for all $a_r, b_s$ in the unit disc. Clearly
\begin{eqnarray}\label{130}
{\cal C}(\theta)\le g(\theta)\le ||\theta||_1;\;\;\; {\cal Q}(\theta)\le ||\theta||_1.
\end{eqnarray}
$G_d$ is the set of matrices $\theta$ with $g(\theta)\le 1$.
Any matrix $\theta$, after normalisation with $g(\theta)$ gives
$\frac{\theta}{g(\theta)}\in G_d$ and can be used in the Grothendieck formalism.

\item[(2)]
$g^\prime (\theta)$ is the supremum of the set of the values  ${\cal C}(\theta)$ in Eq.(\ref{89}) for all $a_r, b_s$ such that
\begin{eqnarray}\label{11A}
\sum_r|a_r|^2\le d;\;\;\;\sum_s|b_s|^2\le d.
\end{eqnarray}
$G_d^\prime$ is the set of matrices $\theta$ with $g^\prime(\theta)\le 1$.
\end{itemize}
\end{definition}

It has been proved \cite{VOU1} that:
\begin{itemize}
\item
\begin{eqnarray}\label{73}
 g(\theta)\le g^\prime (\theta)=d{\mathfrak s}_{\rm max}
\end{eqnarray}
where ${\mathfrak s}_{\rm max}$ is the maximum singular value of the matrix $\theta$.
Therefore $G_d^\prime\subset G_d$.
\item
A necessary (but not sufficient) condition for ${\cal Q}(\theta)>1$ is that $\theta\in G_d\setminus G_d^\prime$.
This occurs only if Eq(\ref{73}) holds as a strict inequality, so that there is a window
\begin{eqnarray}
 \frac{1}{d{\mathfrak s}_{\rm max}}<\lambda<\frac{1}{g(\theta)}.
\end{eqnarray}
Only in this case ${\cal Q}(\lambda \theta)$ {\bf might be} (the condition is not sufficient) greater than one, for some $\lambda$ in the above window.

It is easily seen that if $\theta\in G_d^\prime$ then $U\theta U^\dagger \in G_d^\prime$, where $U$ is any unitary transformation.
In contrast, if $\theta\in G_d\setminus G_d^\prime$ then $U\theta U^\dagger$ might {\bf not} belong in $G_d\setminus G_d^\prime$.
Unitary transformations play a fundamental role in isolated systems. The fact that $\theta \in G_d\setminus G_d^\prime$ is a necessary condition for ${\cal Q}>1$, indicates that
${\cal Q}>1$ can be linked to open systems and irreversible phenomena.

\item
Another necessary (but not sufficient) condition for $\theta\in G_d$ to give ${\cal Q}>1$,
 is that $g(\theta)< ||\theta||_1$ (strict inequality in Eq.(\ref{130})). We combine this with Eq.(\ref{73}) as a strict inequality, and we get the following strict inequality
as a necessary (but not sufficient) condition for ${\cal Q}>1$:
\begin{eqnarray}\label{155}
g(\theta)<\min\left(d{\mathfrak s}_{\rm max}, ||\theta||_1 |\right ).
\end{eqnarray}
\item
If $U$ is a unitary transformation then in general
\begin{eqnarray}\label{17}
{\cal N}(UVU^\dagger)\ne {\cal N}(V);\;\;\;g(U\theta U^\dagger)\ne g(\theta);\;\;\;g^\prime(U\theta U^\dagger)= g^\prime (\theta).
\end{eqnarray}
\end{itemize}

We next rewrite the Grothendieck inequality for {\bf arbitrary} matrices as
 \begin{eqnarray}\label{456}
{\cal Q}\left (\frac{\theta}{g(\theta)}\right )=\left |{\rm Tr}\left (\frac{W^\dagger}{{\cal N}(W)}\frac{\theta}{g(\theta)}\frac{V}{{\cal N}(V)}\right )\right |\le k_G.
\end{eqnarray} 
We note that:
\begin{itemize}
\item
An arbitrary matrix $\theta$ is normalised so that the assumption ${\cal C}\left (\frac{\theta}{g(\theta)}\right )\le 1$ holds.
The role of $g(\theta)$ is to keep the values of the quadratic form ${\cal C}$ with the fraction $\frac{\theta}{g(\theta)}$, bounded.
\item
Arbitrary matrices $V,W$ are normalised so that $\frac{V}{{\cal N}(V)}, \frac{W}{{\cal N}(W)}$ are rescaling matrices in ${\cal S}_d$.
\item
The ${\rm Tr}(W^\dagger \theta V)$ is invariant under unitary transformations, but the $g(\theta)$, ${\cal N}(V)$, ${\cal N }(W)$ are not invariant under unitary transformations.
Therefore ${\cal Q}$  {\bf is not invariant under unitary transformations}, and this makes the formalism suitable for open systems and irreversible phenomena.
Unitary transformations are a feature of isolated quantum system. 
In open systems we have no invariance under unitary transformations, and in this paper we use rescaling transformations.
\end{itemize}

In the following we show that for  a pure state in an isolated system ($\theta=\ket{f}\bra{f}$), the corresponding ${\cal Q}\le 1$. 
For some types of mixed states we also get ${\cal Q}\le 1$. Later we show that in an open system ${\cal Q}$ might take values in the region $(1,k_G)$.

\begin{proposition}\label{L50}
In an isolated system $\Sigma(d)$ we consider the following:
\begin{itemize}
\item[(1)]
The density matrix $\theta_{rs}=f_rf_s^*$ for a quantum state $\ket{f}$ with components $f_r$ in some orthonormal basis ($\sum _r|f_r|^2=1$). Then
\begin{eqnarray}\label{pure}
g(\theta)=||f_rf_s^*||_1\ge 1;\;\;\;{\cal Q}\left(\frac{\theta}{g(\theta)}\right)\le 1.
\end{eqnarray}

\item[(2)]
A diagonal density matrix
\begin{eqnarray}
\theta={\rm diag}(p_0,\cdots,p_{d-1});\;\;\;\sum_rp_r=1;\;\;\;p_r\ge 0.
\end{eqnarray}
Then 
\begin{eqnarray}\label{29}
g(\theta)= ||\theta ||_1=1;\;\;\;{\cal Q}(\theta)\le 1.
\end{eqnarray}
\item[(3)]
In Eq.(\ref{456}) we take $\theta$ to be a density matrix and $V=W=\exp(-i{\mathfrak H}t)$ to be a time evolution operator, where ${\mathfrak H}$ is a Hamiltonian.
Then $Q$ does not depend on the time $t$.

\end{itemize}
\end{proposition}
\begin{proof}
\begin{itemize}
\item[(1)]
We first point out that the inequality
\begin{eqnarray}\label{49}
\left |\sum _r \zeta_r\right |\le \sum _r|\zeta_r|
 \end{eqnarray}
becomes equality, only if $\arg(\zeta_r)$ is a constant $\phi$ (independent of the index $r$).

We start from
\begin{eqnarray}
{\cal C}(\theta)=|\sum _{r,s}f_rf_s^*a_rb_s|;\;\;\;|a_r|\le 1\;\;\;|b_s|\le 1.
\end{eqnarray}
We choose $a_r,b_s$ such that $f_ra_r=|f_r|$ and $f_s^*b_s=|f_s|$ and we get the maximum value of all ${\cal C}(\theta)$ (for all $|a_r|\le 1$ and $|b_s|\le 1$), which is by definition $g(\theta)$:
\begin{eqnarray}\label{32}
g(\theta)=\left (\sum _{r}|f_r|\right )^2=||f_rf^*_s||_1\ge \sum _{r}|f_r|^2=1.
\end{eqnarray}
Therefore the strict inequality in Eq.(\ref{155}) which is a necessary condition for ${\cal Q}>1$ is not satisfied, and we conclude that ${\cal Q}\le 1$.

We also give a second (direct) proof that ${\cal Q}\le 1$:
\begin{eqnarray}
\left |\sum_{r,s} \theta_{r,s}\lambda_r\mu_s\bra{u_r}v_s\rangle \right |\le \sum_{r,s} |\theta_{r,s}|=||f_rf^*_s||_1.
\end{eqnarray}
Since we proved above that $g(\theta)=||f_rf^*_s||_1$, it follows that ${\cal Q}\left(\frac{\theta}{g(\theta)}\right )\le 1$.

\item[(2)]
We get
\begin{eqnarray}
{\cal C}(\theta)=|\sum _{r}p_ra_rb_r|;\;\;\;|a_r|\le 1\;\;\;|b_s|\le 1.
\end{eqnarray}
We choose $a_r=b_r=1$ and we easily prove that $g(\theta)= ||\theta ||_1=1$.
Therefore the strict inequality in Eq.(\ref{155}) which is a necessary condition for ${\cal Q}>1$ is not satisfied, and we conclude that ${\cal Q}\le 1$.

We also give a second (direct) proof that ${\cal Q}\le 1$.
We use for ${\cal Q}$ the original expression in Eq.(\ref{34}), and we get
\begin{eqnarray}
{\cal Q}(\theta)=\left |\sum_{r}p_r\lambda_r\mu_r\bra{u_r}v_r\rangle \right |\le \sum_{r}p_r\lambda_r\mu_r|\bra{u_r}v_r\rangle |\le 1.
\end{eqnarray}
\item[(3)]
For unitary operators $V,W$ the ${\cal N}(V)={\cal N}(W)=1$.
Also ${\rm Tr}(W^\dagger \theta V)={\rm Tr}(\theta)$ and therefore ${\cal Q}$ does not depend on $t$.

\end{itemize}
\end{proof}

\subsection{${\cal Q}$ is a generalisation of well known physical quantities}\label{sec46}

In Eq.(\ref{456}) we put $W={\bf 1}$ and $V$ a unitary operator (then ${\cal N}(V)={\cal N}(W)=1$), and we get
 \begin{eqnarray}\label{CC}
{\cal Q}=\frac{ |{\rm Tr}(\theta V )|}{g(\theta)}
\end{eqnarray} 
Many important physical quantities are given by such formula.
Below $\ket{f}$ is a pure state with components $f_r$.
\begin{itemize}
\item
Here $\theta=\ket{f}\bra{f}$, $V={\mathfrak D}(a,b, c)$ are the displacement operators in Eq.(\ref{D}) (they are unitary matrices and therefore rescaling matrices) and $W={\bf 1}_d$. 
Then we use Eq.(\ref{CC}) (with $g(\theta)$ given in Eq.(\ref{32})), and we get
\begin{eqnarray}\label{W1}
{\cal Q}(a,b)=\frac{|\bra{f}{\mathfrak D}(a,b, c)\ket{f}|}{||f_rf^*_s||_1}=\frac{|{\widetilde{\cal W}}(a,b)|}{||f_rf^*_s||_1},
\end{eqnarray}
where ${\widetilde{\cal W}}(\alpha, \beta)$ is the Weyl function of the state $\ket{f}$. 
\item
Here $\theta=\ket{f}\bra{f}$, $V={\mathfrak P}(a,b)$ are the displaced parity operators (defined in e.g., \cite{VB}) and $W={\bf 1}_d$. 
Then we use Eq.(\ref{CC}) (with $g(\theta)$ given in Eq.(\ref{32})), and we get
\begin{eqnarray}\label{W2}
{\cal Q}(a,b)=\frac{|\bra{f}{\mathfrak P}(a,b)\ket{f}|}{||f_rf^*_s||_1}=\frac{|{\cal W}(a,b)|}{||f_rf^*_s||_1},
\end{eqnarray}
 where ${\cal W}(a,b)$ is the Wigner function of the state $\ket{f}$. 
\item
Here $\theta=\ket{f}\bra{f}$, $W={\bf 1}_d$
and $V=U\ket{X;\nu}\bra{X;\nu}U^\dagger$ where $U$ is a unitary matrix. We first prove that $V$ is a rescaling matrix. Indeed
$V_{rs}=U_{r\nu}U^*_{s\nu}$ and
\begin{eqnarray}
\sum _s|V_{rs}|^2=|U_{r\nu}|^2\sum _s|U_{s\nu}|^2=|U_{r\nu}|^2\le 1.
\end{eqnarray}
Then we use Eq.(\ref{CC}) (with $g(\theta)$ given in Eq.(\ref{32})), and we get
\begin{eqnarray}
{\cal Q}=\frac{|\bra{f}U\ket{X;\nu}|^2}{||f_rf^*_s||_1} .
\end{eqnarray}

If $U$ are symplectic transformations, then $|\bra{f}U\ket{X;\nu}|^2$ are measurable quantities in tomography experiments, 
from which we can construct the Wigner function with inverse Radon transform (e.g., \cite{VB}).
\item
Here $\theta_\nu=\ket{f}\bra{X;\nu}$, $W={\bf 1}_d$ and  $V=\exp(i{\mathfrak H} t)$ is the time evolution operator with Hamiltonian ${\mathfrak H}$.
In this case $\theta_{rs}=f_r\delta_{\nu s}$, and we get 
\begin{eqnarray}
{\cal C}(\theta_\nu)=\left |\sum _r f_ra_rb_\nu \right |;\;\;\;|a_r|\le 1;\;\;\;|b_\nu|\le 1.
\end{eqnarray}
Therefore for $b_\nu=1$ and $a_r$ such that $f_ra_r=|f_r|$ we get the supremum of all ${\cal C}(\theta_\nu)$, which  is $g(\theta_\nu)=\sum _r |f_r|$. Using this we get
\begin{eqnarray}
{\cal Q}_\nu(t)=\frac{|\bra{X;\nu}f(t)\rangle |}{\sum _r |f_r|} ;\;\;\;\ket{f(t)}=\exp(i{\mathfrak H} t)\ket{f}
\end{eqnarray}
In this case ${\cal Q}_\nu(t)$ is related to the components of the time evolution $\ket{f(t)}$ of $\ket{f}$.

\end{itemize}

\section{Semi-unitary $d\times 2d$ matrices: coherent states in disguise}

We introduce $2d$ coherent states in the Hilbert space $H(d)$, which we write as columns in  $d\times 2d$ matrices that we call semi-unitary matrices.
They will be used in section 5, to define Bargmann-like representations, that embed a Hilbert space $H(d)$ into a larger Hilbert space $H(2d)$.
Physically, an open system $\Sigma(d)$ will become a subsystem of a larger isolated system $\Sigma(2d)$, and this will be linked to probability currents and the Grothendieck formalism.

In $H(d)$ we consider the $d$ vectors
\begin{eqnarray}\label{100}
&&\ket{z}=\frac{1}{\sqrt 2}\begin{pmatrix}
z^*&1&0&\cdots&0
\end{pmatrix}^\dagger;\;\;\;|z|=1\nonumber\\
&&X\ket{z}=\frac{1}{\sqrt 2}\begin{pmatrix}
1&0&\cdots&0&z^*
\end{pmatrix}^\dagger\nonumber\\
&&\vdots\nonumber\\
&&X^{d-1}\ket{z}=\frac{1}{\sqrt 2}\begin{pmatrix}
0&z^*&1&0&\cdots&0
\end{pmatrix}^\dagger.
\end{eqnarray}
Here $z$ is a complex constant with $|z|=1$.
The vector $\ket{z}$ is a `fiducial vector' and acting with the elements of ${\cal G}_d$ we get the $d$ vectors in Eq.(\ref{100}).

We define the $d\times d$ matrix $A(z)$ that has as columns these vectors (times $\frac{1}{\sqrt 2}$), and also the `parity operator' $\varpi$:
\begin{eqnarray}
A(z)=\frac{1}{2}\begin{pmatrix}
z&1&0&\cdots&0\\
1&0&0&\cdots&z\\
\vdots&\vdots&\vdots&\ddots&\vdots\\
0&0&z&\cdots&0\\
0&z&1&\cdots&0
\end{pmatrix};\;\;\;
\varpi=\begin{pmatrix}
0&1&0&\cdots&0\\
1&0&0&\cdots&0\\
\vdots&\vdots&\vdots&\ddots&\vdots\\
0&0&0&\cdots&0\\
0&0&1&\cdots&0
\end{pmatrix};\;\;\;|z|=1.
\end{eqnarray}
Then 
\begin{eqnarray}
&&\varpi^2=(X\varpi)^2={\bf 1}_d;\;\;\;A(z)=\frac{1}{2}({\bf 1}_d+zX)\varpi\nonumber\\
&&X\varpi=\varpi X^\dagger;\;\;\;[A(z)]^\dagger=A(z^*);\;\;\;A(z)+A(-z)=\varpi.
\end{eqnarray}
Both $\varpi$ and $X\varpi$ are parity operator in the sense that $\varpi^2=(X\varpi)^2={\bf 1}_d$.

We now easily prove the property (where $|z_1|=|z_2|=1$):
\begin{eqnarray}\label{31}
A(z_1)A(z_2^*)+A(-z_1)A(-z_2^*)=\frac{1+z_1z_2^*}{2}{\bf 1}_d.
\end{eqnarray}
Special cases of this are the relations
\begin{eqnarray}\label{AA}
&&\kappa(z)+\kappa(-z)={\bf 1}_d;\;\;\;\lambda(z)+\lambda(-z)=0\nonumber\\
&&\kappa(z)=A(z^*)A(z);\;\;\;\lambda=A(z^*)A(-z).
\end{eqnarray}
Also we find that the diagonal elements of $\kappa(z)$ are equal to $\frac{1}{2}$:
\begin{eqnarray}\label{dia}
[\kappa(z)]_{ii}=\frac{1}{2}.
\end{eqnarray}

We next consider the $d\times 2d$ matrix (written in the `block matrix notation'):
\begin{eqnarray}\label{65}
M(z)=
\begin{pmatrix}
A(z)&A(-z)
\end{pmatrix}=
\begin{pmatrix}
A(z)&\varpi-A(z)
\end{pmatrix}
\end{eqnarray}
Here the `row index' takes values from $0$ to $d-1$, and the `column index' from $0$ to $2d-1$.
We call them semi-unitary matrices for reasons discussed below.
Also in section \ref{sec20} we show that their columns are some type of coherent states in the Hilbert space $H(d)$.
\begin{lemma}\label{LL1}
If $|z_1|=|z_2|=1$, then
\begin{itemize}
\item[(1)]
\begin{eqnarray}\label{67}
M(z_1)[M(z_2)]^\dagger=\frac{1+z_1z_2^*}{2}{\bf 1}_d.
\end{eqnarray}
\item[(2)]
We define the $2d\times 2d$ matrix $\Pi(z_1,z_2)$ (for $z_1\ne -z_2$) as
\begin{eqnarray}\label{321}
\Pi(z_1,z_2)=\frac{2}{1+z_1^*z_2}[M(z_1)]^\dagger M(z_2).
\end{eqnarray}
Then
\begin{eqnarray}
&&[\Pi(z_1,z_2)]^2=\Pi(z_1,z_2);\;\;\;[\Pi(z_1,z_2)]^\dagger=\Pi(z_2,z_1);\;\;\;{\rm Tr}[\Pi(z_2,z_1)]=d.
\end{eqnarray}

$\Pi(z_1,z_2)$ is an idempotent matrix (projector) which in general is not Hermitian, unless $z_1=z_2$ (see below). $\Pi(z_1,z_2)$ is not an orthogonal projector.
The physical significance of $\Pi(z_1,z_2)$ is discussed below (Eqs(\ref{E1}), (\ref{E2})).
\end{itemize}
\end{lemma}
\begin{proof}
\begin{itemize}
\item[(1)]
Using Eq(\ref{31}) and the block matrix notation we get
\begin{eqnarray}\label{58}
M(z_1)[M(z_2)]^\dagger=A(z_1)[A(z_2)]^\dagger+A(-z_1)[A(-z_2)]^\dagger=\frac{1+z_1z_2^*}{2}{\bf 1}_d.
\end{eqnarray}
\item[(2)]
Using Eq.(\ref{67}) we get
\begin{eqnarray}
[M(z_1)]^\dagger M(z_2)[M(z_1)]^\dagger M(z_2)=\frac{1+z_1^*z_2}{2}[M(z_1)]^\dagger M(z_2).
\end{eqnarray}
Therefore the $\Pi(z_1,z_2)=\frac{2}{1+z_1^*z_2}[M(z_1)]^\dagger M(z_2)$ is an idempotent matrix (projector).

We easily prove $[\Pi(z_1,z_2)]^\dagger=\Pi(z_2,z_1)$ and therefore
in general this matrix is not Hermitian, i.e., it is not an orthogonal projector.

From Eq.(\ref{58}) follows that 
\begin{eqnarray}
{\rm Tr}[M(z_1)[M(z_2)]^\dagger]=d\frac{1+z_1z_2^*}{2}\Rightarrow {\rm Tr}[[M(z_2)]^\dagger M(z_1)]=d\frac{1+z_1z_2^*}{2}.
\end{eqnarray}
From this and Eq.(\ref{321}) follows that ${\rm Tr}[\Pi(z_2,z_1)]=d$.

\end{itemize}
\end{proof}
From this lemma follows  immediately the corollary below.
\begin{cor}
If $|z|=|z_1|=|z_2|=1$ then
\begin{itemize}
\item[(1)]
\begin{eqnarray}\label{30}
M(z)[M(z)]^\dagger={\bf 1}_d;\;\;\;M(-z)[M(z)]^\dagger=0.
\end{eqnarray}
\item[(2)]
The $2d\times 2d$ matrix 
\begin{eqnarray}
\Pi(z,z)\equiv \Pi(z)=[M(z)]^\dagger M(z);\;\;\;[\Pi(z)]^2=\Pi(z);\;\;\;{\rm Tr}[\Pi(z)]=d;\;\;\;{\rm rank}[\Pi(z)]=d.
\end{eqnarray}
is an orthogonal projector (Hermitian idempotent matrix).
$\Pi(z)$ has $d$ eigenvalues equal to $1$, and $d$ eigenvalues equal to $0$.
An example of $\Pi(z)$ with $2d=6$ is given later in Eq.(\ref{P12}).
\item[(3)]
\begin{eqnarray}\label{15}
\Pi(z_1)\Pi(z_2)=\frac{2+z_1^*z_2+z_1z_2^*}{4}\Pi(z_1,z_2).
\end{eqnarray}
\item[(4)]
\begin{eqnarray}
&&\Pi(z_1)\Pi(z_1,z_2)=\Pi(z_1,z_2);\;\;\;\Pi(z_1,z_2)\Pi(z_1)=\Pi(z_1)\nonumber\\
&&\Pi(z_2)\Pi(z_1,z_2)=\Pi(z_2);\;\;\;\Pi(z_1,z_2)\Pi(z_2)=\Pi(z_1,z_2).
\end{eqnarray}
\end{itemize}
\end{cor}

Since $M(z)[M(z)]^\dagger={\bf 1}_d$ and $[M(z)]^\dagger M(z)=\Pi(z)$ is projector (which is a unit matrix within the space that it projects into), we call 
the matrices $M(z)$ semi-unitary. We see below that they can be used in a similar way to unitary matrices.

\begin{proposition}
The $2d\times 2d$ orthogonal projector $\Pi(z)$ has the following properties:
\begin{itemize}
\item[(1)]
\begin{eqnarray}\label{66}
\Pi(z)+\Pi(-z)={\bf 1}_{2d};\;\;\;\Pi(z)\Pi(-z)=0;\;\;\;|z|=1.
\end{eqnarray}
Also
\begin{eqnarray}\label{300}
M(z)\Pi(z)=M(z);\;\;\;M(-z)\Pi(z)=0.
\end{eqnarray}
\item[(2)]
All eigenvectors of $\Pi(z)$ span a $2d$-dimensional space $H(2d)$.
The eigenvectors of $\Pi(z)$ corresponding to the eigenvalue one, span the $d$-dimensional space $H(d)$.
The eigenvectors of $\Pi(z)$ corresponding to the eigenvalues zero, span a $d$-dimensional space  $H_{\rm null}(d)$
(which mathematically is a null space and physically `external world'). Then
\begin{eqnarray}\label{780}
H(d)\oplus H_{\rm null}(d)=H(2d).
\end{eqnarray}
The projector $\Pi(-z)$ projects into $H_{\rm null}(d)$.

\item[(3)]
$\Pi(z)$ is a rescaling matrix with diagonal elements equal to $\frac{1}{2}$ and
\begin{eqnarray}\label{6BA}
{\cal N}[\Pi(z)]=\frac{1}{\sqrt{2}};\;\;\;|z|=1.
\end{eqnarray}
\item[(4)]
The following strict inequality holds for almost all $z$ (except from a finite number of values of $z$ on the unit circle):
\begin{eqnarray}\label{400}
g[\Pi(z)]<2d.
\end{eqnarray}

\end{itemize}
\end{proposition}
\begin{proof}
\mbox{}
\begin{itemize}
\item[(1)]
We first write $\Pi(z)$ in the block matrix notation as
\begin{eqnarray}\label{333}
\Pi(z)=\begin{pmatrix}
\kappa(z)&\lambda(z)\\
\lambda(-z)&\kappa(-z)
\end{pmatrix}=
\begin{pmatrix}
\kappa(z)&\lambda(z)\\
-\lambda(z)&{\bf 1}_d-\kappa(z)
\end{pmatrix}.
\end{eqnarray}
and then using Eq.(\ref{AA}) we prove that $\Pi(z)+\Pi(-z)={\bf 1}_{2d}$.

We prove that $\Pi(z)\Pi(-z)=0$ using Eq.(\ref{15}).
Eq.(\ref{300}) follows immediately from Eq.(\ref{30}).
\item[(2)]
This is rather straightforward.
Eq.(\ref{66}) shows that the projector $\Pi(-z)$ projects into $H_{\rm null}(d)$.

\item[(3)]

Direct multiplication of $[M(z)]^\dagger$ with $M(z)$ shows that every row in $\Pi(z)$ has:
\begin{itemize}
\item
 diagonal element $[{\Pi}(z)]_{ii}=\frac{1}{2}$ (related to this is Eq.(\ref{dia}))
 \item
four other elements $\pm \frac{z}{4}$ or $\pm \frac{z^*}{4}$ (their absolute values are $\frac{1}{4}$)
\item
all the other $2d-5$ elements are zero.
\end{itemize}  
An example is given in Eq.(\ref{P12}) below.
From this follows Eq.(\ref{6BA}).

\item[(4)]

We get
\begin{eqnarray}
{\cal C}[\Pi(z)]=\left |\vec{a}\cdot [\Pi (z)\vec{b}]\right |= \left |\vec{a}\cdot\vec{\beta}\right |;\;\;\;\vec{\beta}=\Pi(z)\vec{b}.
\end{eqnarray}
Here $\vec{a}$ is a vector (row) that contains the $a_r$ (with $|a_r|\le1$), and 
$\vec{b}$ is a vector (column) that contains the elements $b_r$ (with $|b_r|\le 1$).
Clearly $|\vec{a}|^2\le 2d$. Also $|\vec{b}|^2\le 2d$ and since $\vec{\beta}$ is a projection of $\vec{b}$, we also have  $|\vec{\beta}|^2\le 2d$. Then
\begin{eqnarray}
{\cal C}[\Pi(z)]=\left |\vec{a}\cdot\vec{\beta}\right |\le |\vec{a}||\vec{\beta}|\le 2d.
\end{eqnarray}
For ${\cal C}[\Pi(z)]=2d$ we need  $ |\vec{a}|=|\vec{\beta}|=\sqrt{ 2d}$ and $\vec{a}=\vec{\beta}$, which implies that we need to have $a_r=\beta_r$ with absolute values $|a_r|=|\beta_r|=1$ for all $r$.
We next show that for at least some of the $\beta_r$, there exist infinitesimals $\epsilon_r>0$ such that  $|\beta_r|<1-\epsilon_r$. Therefore there exist  infinitesimal $\epsilon$ such that ${\cal C}[\Pi(z)]<2d-\epsilon$ which in turn implies the strict inequality in Eq.(\ref{400}).

The $\vec{\beta}=\Pi (z)\vec{b}$ implies that $\Pi(-z)\vec{\beta}=0$, and since ${\rm rank}[\Pi(z)]=d$ this gives $d$ independent equations.
So we can choose $\beta_0,...,\beta_{d-1}$ such that $|\beta_0|=|\beta_{d-1}|=1$, but then the solutions for $\beta_d,...,\beta_{2d-1}$ will not give $|\beta_r|\in (1-\epsilon_r,1)$ for $r=d,...,2d-1$ and for {\bf all}
$z$ on the unit circle ($|z|=1$). 

We give an example where for a {\bf particular value of} $z$, we do get all $|\beta_r|=1$ for all $r$.
In the case $d=3$, the $3\times 6$ matrix $M(z)$ and the $6\times 6$ matrix $\Pi(z)$ are given in Eqs.(\ref{AQ1}),(\ref{P12}) below.
We take $z=i$ and we find
\begin{eqnarray}
\vec{\beta}=\sqrt{2}[M(i)]^\dagger 
\begin{pmatrix}
1&1&1
\end{pmatrix}^\dagger=
\begin{pmatrix}
w&w&w&w^*&w^*&w^*
\end{pmatrix}^\dagger;\;\;\;w=\exp\left(i\frac{\pi}{4}\right)
\end{eqnarray}
For this particular value of $z$, all $|\beta_r|=1$ and $\Pi(-i)\vec{\beta}=0$, and the strict inequality in Eq.(\ref{400}) does not hold. But for a generic $z$, the strict inequality in Eq.(\ref{400}) does hold.
\end{itemize}
\end{proof}

\subsection{The columns of $M(z)$ are coherent states in $H(d)$}\label{sec20}

 We consider the columns of $M(z)$, i.e., the  vectors in Eq.(\ref{100}) with both $z$ and $-z$:
\begin{eqnarray}\label{reso1}
{\mathfrak S}_d=\{\ket{z},...,X^{d-1}\ket{z}, \ket{-z},...,X^{d-1}\ket{-z}\}.
\end{eqnarray}

The relation  $M(z)[M(z)]^\dagger={\bf 1}_d$  is a resolution of the identity in terms of them, and can be written as 
\begin{eqnarray}\label{reso}
\frac{1}{2}\left [\ket{z}\bra{z}+...+X^{d-1}\ket{z}\bra{z}(X^\dagger)^{d-1}+\ket{-z}\bra{-z}+...+X^{d-1}\ket{-z}\bra{-z}(X^\dagger)^{d-1}\right ]={\bf 1}_d.
\end{eqnarray}
We note here that the $d$ vectors $\ket{z},X\ket{z},...,X^{d-1}\ket{z}$ span $H(d)$, but do not resolve the identity 
\begin{eqnarray}
\ket{z}\bra{z}+...+X^{d-1}\ket{z}\bra{z}(X^\dagger)^{d-1}\ne {\bf 1}_d.
\end{eqnarray}
The $2d$ states in the set ${\mathfrak S}_d$ are coherent states in the sense that that they resolve the identity, and are  associated with a group (the group ${\cal G}_d$).
We note however two differences in comparison with other known sets of coherent states:
\begin{itemize}
\item
The action of the group ${\cal G}_d$ on these states is not transitive. We have two orbits:
\begin{eqnarray}
&&\ket{z}\xrightarrow {X}  X\ket{z}\xrightarrow {X}...\xrightarrow {X}X^{d-1}\ket{z}\xrightarrow {X}  \ket{z}\nonumber\\
&&\ket{-z}\xrightarrow {X}  X\ket{-z}\xrightarrow {X}...\xrightarrow {X}X^{d-1}\ket{-z}\xrightarrow {X}  \ket{-z}
\end{eqnarray}
\item
Some of these states are orthogonal to each other (the states $X^r\ket{z}$ and  $X^r\ket{-z}$ are orthogonal for all $r$).
\end{itemize}
States similar to these (with dimension $d=3$ and $d=4$) have been studied in ref.\cite{VOU2} with the name ultra-quantum coherent states (for reasons discussed there). 

\section{$z$-Bargmann representations for  open systems}\label{sec10}

In this section each vector in $H(d)$ is represented with $2d$ coordinates, the overlaps of the vector with the $2d$ coherent states in Eq.(\ref{reso1}).
This is a Bargmann-like representation, 
but  in comparison  with the one for the harmonic oscillator, we mention two important differences:
\begin{itemize}
\item
The Bargmann representation for the harmonic oscillator\cite{BA} does not enlarge the Hilbert space.
In the present paper the enlargement of the Hilbert space in Eq.(\ref{780}) is crucial.
$H(d)$ is embedded into a bigger space $H(2d)$.
From a physical point of view, an open system $\Sigma(d)$ becomes a subsystem of a larger isolated system $\Sigma(2d)$ (the `full universe').
Unitary transformations take place in the isolated system $\Sigma(2d)$ (e.g., time evolution), and they are projected into rescaling transformations in the open system $\Sigma(d)$ (proposition \ref{L1} below).

\item
One of the merits of the Bargmann representation for the harmonic oscillator, is analyticity and the use of the theory of analytic functions in a quantum context.
In our discrete context there is no analyticity (but some kind of analyticity in a discrete context is discussed in \cite{BA1,BA2}).
\end{itemize}

The $z$-Bargmann representation introduces naturally $H_{\rm null}(d)$ in Eq.(\ref{780}), which mathematically is a null space and physically is an `external world'.
This makes naturally the system $\Sigma(d)$ an open system.

We compare and contrast the tensor product approach to the environment, with the direct sum approach used in this paper.
The former involves a Hamiltonian that describes the interaction between environment and the system, with various coupling constants that in principle can be found from experimental results.
The latter is an effective formalism which is simpler, and has the parameter $z$ (with $|z|=1$).
Different values of $z$ describe different embeddings of $\Sigma(d)$ into $\Sigma(2d)$.

\begin{remark}
There are other very different areas where there is merit in representing a vector in $H(d)$ with another vector in a larger space:
\begin{itemize}
\item 
Naimark dilation (e.g., \cite{HOL}): A positive operator valued measure (POVM) in a $d$-dimensional Hilbert space can be linked to a projection valued measurement (PVM) in a Hilbert space with larger dimension.
We note however, that this construction involves a tensor product of Hilbert spaces, in contrast to our construction which involves a direct sum of Hilbert spaces.
\item
Thermofield dynamics (e.g., \cite{AA0}): A mixed state is represented by a pure state in a larger space.
\item
Reproducing kernel Hilbert space formalism in the context of artificial intelligence and big data (e.g., \cite{AA1,AA2,AA3}):
Data are separated easier when they are embedded in a larger space.
\end{itemize}
\end{remark}

\subsection{$z$-Bargmann representation of quantum states}

Using the resolution of the identity in Eq.(\ref{reso}) we represent an arbitrary vector $\ket{v}$ in $H(d)$ in terms of $2d$ components as
\begin{eqnarray}
\ket{v}=v_0(z)\ket{z}+...+v_{d-1}(z)X^{d-1}\ket{z}+v_d(z)\ket{-z}+....+v_{2d-1}(z)X^{d-1}\ket{-z},
\end{eqnarray}
where
\begin{eqnarray}\label{13}
&&0\le r\le d-1\;\rightarrow\; v_r(z)=\frac{1}{2}\bra{z}(X^\dagger)^{r}\ket{v}\nonumber\\
&&d\le r\le 2d-1\;\rightarrow\; v_r(z)=\frac{1}{2}\bra{-z}(X^\dagger)^{r-d}\ket{v}.
\end{eqnarray}
Eq.(\ref{13}) can also be written in terms of the matrix $M(z)$:
\begin{eqnarray}\label{78}
v_B(z)=[M(z)]^\dagger v=\begin{pmatrix}
A(z^*)v\\
[\varpi-A(z^*)]v
\end{pmatrix};\;\;\;|z|=1.
\end{eqnarray}
This relates the $d$ components of $\ket{v}$ in the orthonormal basis in Eq.(\ref{1}), with its 
$2d$ $z$-Bargmann components (the index B indicates Bargmann).
Not every $2d$-tuple is a $z$-Bargmann representation of a vector in $H(d)$.
Only $2d$-tuples which are eigenstates of $\Pi(z)$ with eigenvalue one
\begin{eqnarray}\label{W3}
\Pi(z)v_B(z)=v_B(z),
\end{eqnarray}
are $z$-Bargmann representations of some vector in $H(d)$.
This is proved using the fact that $M(z)[M(z)]^\dagger={\bf 1}_d$.

\subsection{$z$-Bargmann representation of matrices}

For an arbitrary $d\times d$ matrix $T$ (e.g, a density matrix or a transformation), the $z$-Bargmann representation is  the $2d\times 2d$ matrix 
\begin{eqnarray}\label{93}
T_B(z)=[M(z)]^\dagger TM(z);\;\;\;|z|=1.
\end{eqnarray}
For example, the density matrix  $\rho=\frac{1}{d}{\bf 1}_d$ (the mixed state with maximum entropy) becomes
\begin{eqnarray}\label{93A}
\rho_B(z)=\frac{1}{d}[M(z)]^\dagger {\bf 1}_dM(z)=\frac{1}{d}\Pi(z).
\end{eqnarray}
Not every $2d\times 2d$ matrix is a $z$-Bargmann representation of a $d\times d$ matrix.
Only those which obey the relation
\begin{eqnarray}\label{W30}
\Pi(z)T_B(z)\Pi(z)=T_B(z)
\end{eqnarray}
are $z$-Bargmann representations of a $d\times d$ matrix.

We note that:
\begin{itemize}
\item
The matrix $T_B(z)$ cannot be equal to ${\bf 1}_{2d}$ because this does not obey Eq.(\ref{W30}).

\item
Unitary $d\times d$ matrices $TT^\dagger ={\bf 1}_d$ are represented with 
$2d\times 2d$ matrices such that $T_B(z)[T_B(z)]^\dagger=\Pi(z)$. Indeed in this case
\begin{eqnarray}
T_B(z)[T_B(z)]^\dagger=[M(z)]^\dagger TM(z)[M(z)]^\dagger T^\dagger M(z)=[M(z)]^\dagger M(z)=\Pi(z).
\end{eqnarray}
\item
$T_B(z)$ cannot satisfy the $T_B(z)[T_B(z)]^\dagger={\bf 1}_{2d}$, i.e., the $2d\times 2d$ matrix $T_B(z)$ cannot be unitary.
This would require $\Pi(z)T_B(z)\Pi(z)[T_B(z)]^\dagger \Pi(z)={\bf 1}_{2d}$ which is not possible (if we multiply by $\Pi(-z)$ both sides we get that $0$ is equal to $\Pi(-z)$).

\end{itemize}

\subsection{Standard representation vs. $z$-Bargmann representation}\label{sec45}

We have two representations of the same system. The `standard representation'   
represents vectors with $d$ components using an orthonormal basis in $H(d)$, and the $z$-Bargmann representation represents vectors with $2d$ components.
We expect that many basic quantities (scalar product, etc) are the same in both  representations, and we show this explicitly.

\begin{proposition}\label{pro11}
\mbox{}
\begin{itemize}
\item[(1)]
The Bargmann transformation in Eq.(\ref{78}) preserves the scalar product:
\begin{eqnarray}
[v_B(z)]^\dagger u_B(z)=v^\dagger u.
\end{eqnarray}
\item[(2)]
$d$ of the eigenvalues of the $2d\times 2d$ matrix $T_B(z)=[M(z)]^\dagger TM(z)$ are the same as the eigenvalues of the $d\times d$ matrix $T$, and the other $d$ are equal to zero.
\end{itemize}
\end{proposition}
\begin{proof}
\begin{itemize}
\item[(1)]
This follows easily from the fact that $M(z)[M(z)]^\dagger={\bf 1}_d$.
\item[(2)]
If $\lambda$ is an eigenvalue of $T$ and $v$ is the corresponding eigenvector , then
\begin{eqnarray}
Tv=\lambda v\;\Rightarrow\;[M(z)]^\dagger TM(z)[M(z)]^\dagger v=\lambda [M(z)]^\dagger v\;\Rightarrow\;T_B(z)v_B(z)=\lambda v_B(z).
\end{eqnarray}
Therefore $\lambda$ is also an eigenvalue of $T_B(z)$ (with the same multiplicity).
On the other hand
\begin{eqnarray}
Tv=\lambda v\;\Rightarrow\;[M(z)]^\dagger TM(z)[M(-z)]^\dagger v=0\;\Rightarrow\;T_B(z)v_B(-z)=0.
\end{eqnarray}
It follows that the $d$ eigenvectors $v_B(-z)=[M(-z)]^\dagger v$ are eigenvectors of $T_B(z)$ with corresponding eigenvalue zero.
\end{itemize}
\end{proof}

The following properties are easily proved using lemma \ref{LL1} and proposition \ref{pro11}:
\begin{itemize}
\item
If $u=Tv$, its $z$-Bargmann representation is $u_B(z)=T_B(z)v_B(z)$.
For example, the relation ${\bf 1}_d v=v$ becomes $\Pi(z)v_B(z)=v_B(z)$ (Eq.(\ref{W3}) given earlier), which  is the `reproducing kernel' property.

\item
If $T_B$ is the $z$-Bargmann representation of the matrix $T$, then ${\rm Tr}(T_B)={\rm Tr}(T)$.
\item
If $\sigma$ is a $d\times d$ density matrix matrix in $\Sigma(d)$, then $\sigma_B=[M(z)]^\dagger\sigma M(z)$ is a $2d\times 2d$ density matrix in $\Sigma(2d)$.
In this case $\Pi(z)\sigma_B\Pi(z)=\sigma_B$, and the von Neumann entropy of both $\sigma$ and $\sigma_B$ is the same.
For a general $2d\times 2d$ density matrix $\rho$ in $\Sigma(2d)$, we get $\Pi(z)\rho\Pi(z)\ne \rho$ and this leads to a probability current from $\Sigma(d)$ to the external world as discussed  below.
\item
If $T,S$ are two $d\times d$ matrices, the $z$-Bargmann representation of their product $R=TS$ is  $R_B=T_B(z)S_B(z)$.
More generally if there is a relation between $d\times d$ matrices $T, S,...$, then the same relation holds between their 
$2d\times 2d$ $z$-Bargmann representatives
$T_B(z), S_B(z),...$.

\item
$\Pi(z_1,z_2)$ can be used to change representation. Acting on the vector $v_B(z_2)$ in the $z_2$-Bargmann representation,
gives the same vector in the $z_1$-Bargmann representation times a factor (here $z_2\ne -z_1$):
\begin{eqnarray}\label{E1}
\Pi(z_1,z_2)v_B(z_2)=\frac{2}{1+z_1^*z_2}v_B(z_1)
\end{eqnarray}
Also $v_B(z_1)$ are eigenvectors of $\Pi(z_1,z_2)$ :
\begin{eqnarray}
\Pi(z_1,z_2)v_B(z_1)=v_B(z_1).
\end{eqnarray}

An analogous relation for matrices is (here $z_2\ne -z_1$)
\begin{eqnarray}\label{E2}
\Pi(z_1,z_2)T_B(z_2)\Pi(z_2,z_1)=\frac{4}{2+z_1^*z_2+z_1z_2^*}T_B(z_1).
\end{eqnarray}

\end{itemize}

\subsection{Example: the open system $\Sigma(3)$ embedded into the isolated system $\Sigma(6)$}

In the case $d=3$, we get
\begin{eqnarray}\label{AQ1}
M(z)=\frac{1}{2}
\setcounter{MaxMatrixCols}{12}
\begin{pmatrix}
1&z&0&1&-z&0\\
z&0&1&-z&0&1\\
0&1&z&0&1&-z
\end{pmatrix};\;\;\;|z|=1,
\end{eqnarray}
and
\begin{eqnarray}\label{P12}
\Pi(z)=[M(z)]^\dagger M(z)=\frac{1}{4}
\begin{pmatrix}
2&z&z^*&0&-z&z^*\\
z^*&2&z&z^*&0&-z\\
z&z^*&2&-z&z^*&0\\
0&z&-z^*&2&-z&-z^*\\
-z^*&0&z&-z^*&2&-z\\
z&-z^*&0&-z&-z^*&2
\end{pmatrix}.
\end{eqnarray}

As an example we consider the following density matrix, and express it in the $z$-Bargmann representation:
\begin{eqnarray}
\rho=\begin{pmatrix}
1&0&0\\
0&0&0\\
0&0&0
\end{pmatrix}\rightarrow
\rho_B(z)=[M(z)]^\dagger \rho M(z)=\frac{1}{4}\begin{pmatrix}
1&z&0&1&-z&0\\
z^*&1&0&z^*&-1&0\\
0&0&0&0&0&0\\
1&z&0&1&-z&0\\
-z^*&-1&0&-z^*&1&0\\
0&0&0&0&0&0
\end{pmatrix}.
\end{eqnarray}

We use the above example to show that in general
\begin{eqnarray}\label{P0}
g\left \{[M(z)]^\dagger \theta M(z)\right \}\ne g(\theta).
\end{eqnarray}
We show that in the above example $g(\rho)\ne g[\rho_B(1)]$.
The ${\cal C}(\rho)$ (defined in Eq.(\ref{89})) is here ${\cal C}(\rho)=|a_0b_0|$.
Since $g(\rho)$ is the supremum of ${\cal C}(\rho)$ for all $|a_0|\le 1$ and $|b_0|\le 1$, we find $g(\rho)=1$.

The ${\cal C}[\rho_B(z)]$ is the absolute value of a sum with $16$ terms (the non-zero elements in the matrix $\rho_B(z)$ multiplied with $a_rb_s$). We take $z=1$ and
\begin{eqnarray}
a_0=a_1=a_3=-a_4=1;\;\;\;b_0=b_1=b_3=-b_4=1.
\end{eqnarray}
We do not need values for $a_2, b_2, a_5, b_5$.
In this case we find that ${\cal C}[\rho_B(1)]=4$. Therefore $g[\rho_B(1)]$ which is the supremum for all $|a_r|\le 1$ and $|b_r|\le 1$ is $g[\rho_B(1)]\ge 4$.
It is seen that in this example, the two sides in Eq.(\ref{P0}) are different from each other.

Using a different example we also show that in general
\begin{eqnarray}\label{P1}
{\cal N}\left \{[M(z)]^\dagger V M(z)\right \}\ne {\cal N}(V).
\end{eqnarray}
We take $V={\bf 1}_3$ and then  $[M(z)]^\dagger V M(z)=\Pi(z)$.
In this case ${\cal N}\left \{[M(z)]^\dagger V M(z)\right \}=\frac{1}{\sqrt{2}}$ and  ${\cal N}(V)=1$.

\section{Probability currents}\label{prob}

A general density matrix $\rho_0$ in the isolated system $\Sigma(2d)$ (full universe)
evolves in time with the unitary evolution operator $\exp(i{\mathfrak H}t)$, 
where ${\mathfrak H}$ is a $2d\times 2d$ Hermitian matrix describing the Hamiltonian:
\begin{eqnarray}\label{TT}
\rho(t)=\exp(i{\mathfrak H}t)\rho_0\exp(-i{\mathfrak H}t).
\end{eqnarray}
From ${\rm Tr}[\rho(t)]=1$ follows that
\begin{eqnarray}\label{700}
\frac{d}{dt}[{\rm Tr}[\rho(t)]]=0\implies J(t)=\frac{d}{dt}{\rm Tr}[\Pi(z)\rho(t)]=-\frac{d}{dt}{\rm Tr}[\Pi(-z)\rho(t)]
\end{eqnarray}
The $\frac{d}{dt}[{\rm Tr}[\rho(t)]]=0$ can be interpreted as a conservation of probability in $\Sigma(2d)$.
${\rm Tr}[\Pi(z)\rho(t)]$ is the probability that the quantum state described by $\rho(t)$ is in $\Sigma(d)$, and its time derivative is a probability current from the open system
$\Sigma(d)$ to the external world $\Sigma_{\rm ext}(d)$. The sign of the current indicates the direction of the flow.
Positive $J(t)$ indicates amplification in $\Sigma(d)$, and negative $J(t)$ indicates damping in $\Sigma(d)$.
The embedding of $\Sigma(d)$ into the bigger isolated system $\Sigma(2d)$ allows for irreversible phenomena in $\Sigma(d)$, where probability is not conserved. 

If $A$ is time-independent matrix, then
\begin{eqnarray}
\frac{d}{dt}{\rm Tr}[A\rho(t)]=i{\rm Tr}\{[A,{\mathfrak H}]\rho(t)\}
\end{eqnarray}
Therefore the probability current is
\begin{eqnarray}\label{98}
J(t)=i{\rm Tr}\{[\Pi(z), {\mathfrak H}]\rho(t)\}=i{\rm Tr}\{[\Pi(z), {\mathfrak H}]\exp(i{\mathfrak H}t)\rho_0\exp(-i{\mathfrak H}t)\}.
\end{eqnarray}
In a similar way we calculate derivatives of the probability current:
\begin{eqnarray}
&&\frac{dJ(t)}{dt}=-{\rm Tr}\{[ [\Pi(z), {\mathfrak H}],{\mathfrak H}]\rho(t)\}=-{\rm Tr}\{[ [\Pi(z), {\mathfrak H}],{\mathfrak H}]\exp(i{\mathfrak H}t)\rho_0\exp(-i{\mathfrak H}t)\}\nonumber\\
&&\frac{d^2J(t)}{dt^2}=-i{\rm Tr}\{[[[\Pi(z), {\mathfrak H}],{\mathfrak H}],{\mathfrak H}]\rho(t)\}=-i{\rm Tr}\{[[[\Pi(z), {\mathfrak H}],{\mathfrak H}],{\mathfrak H}]\exp(i{\mathfrak H}t)\rho_0\exp(-i{\mathfrak H}t)\}.
\end{eqnarray}

For small $t$,  we can use a truncated Taylor series to evaluate the probability current $J(t)$:
\begin{eqnarray}\label{100}
J(t)=J(0)+\frac{dJ(0)}{dt}t+\frac{d^2J(0)}{dt^2}\frac{t^2}{2!}+...
\end{eqnarray}

\begin{figure}[!htb]
\centering
\includegraphics[width=8cm]{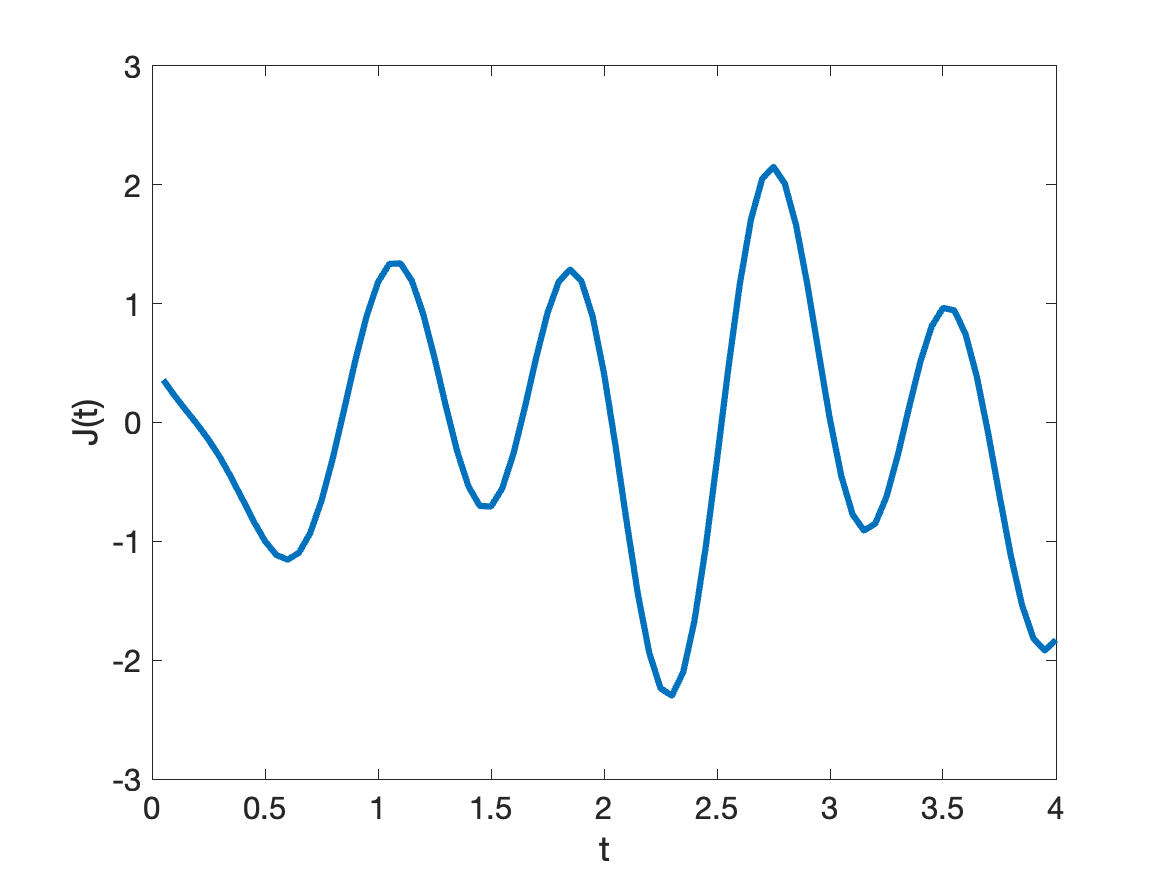}
\caption{The probability current $J(t)$ as a function of the time $t$. At $t=0$ the the system is in the pure state $v_0$ in Eq.(\ref{55}), and evolves unitarily with the Hamiltonian ${\mathfrak H}_1$ in Eq.(\ref{555}).
The parameter $z=\exp\left (i\frac{\pi}{4}\right )$.}
\label{f1}
\end{figure}

\begin{figure}[!htb]
\centering
\includegraphics[width=8cm]{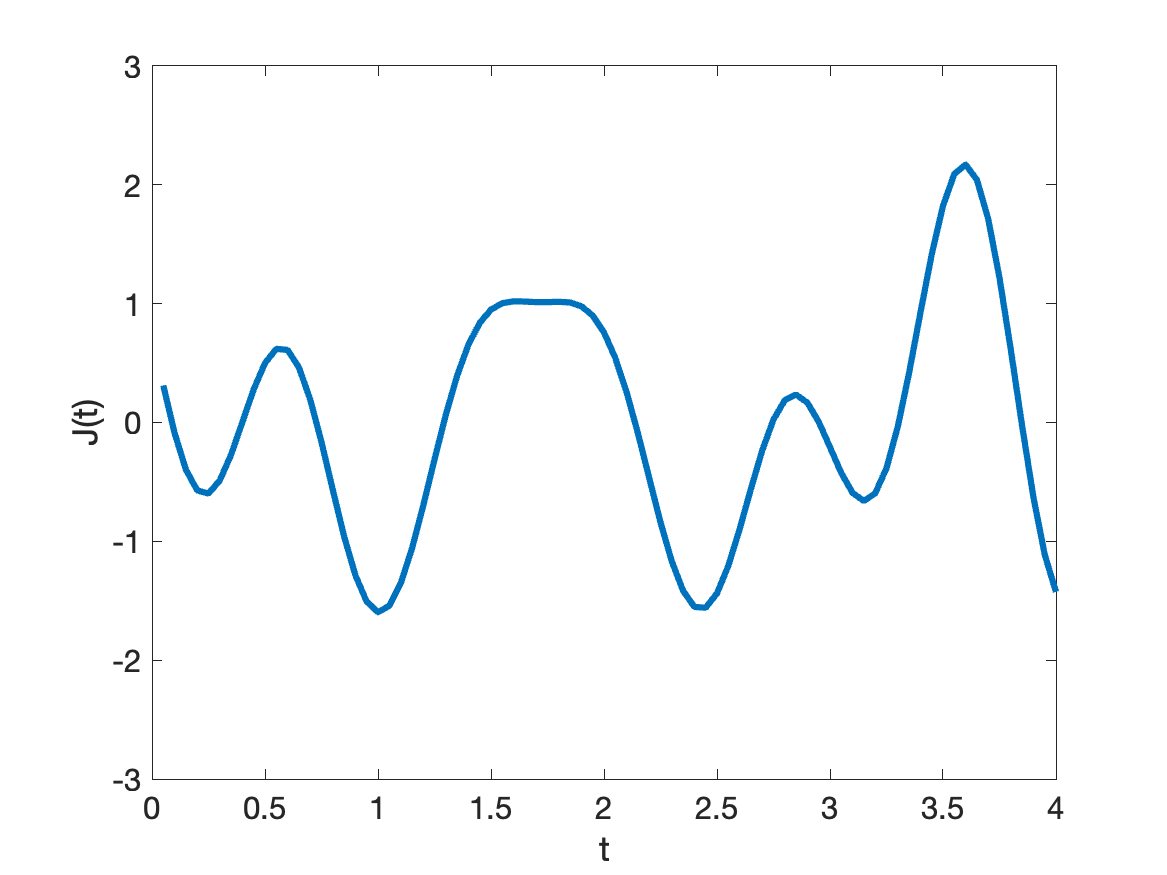}
\caption{The probability current $J(t)$ as a function of the time $t$. At $t=0$ the the system is in the pure state $v_0$ in Eq.(\ref{55}), and evolves unitarily with the Hamiltonian ${\mathfrak H}_1$ in Eq.(\ref{555}).
The parameter $z=\exp\left (i\frac{2\pi}{3}\right )$.}
\label{f2}
\end{figure}

\begin{example}\label{ex10}
In the isolated system (full universe) $\Sigma(6)$ we consider at $t=0$ the pure state 
\begin{eqnarray}\label{55}
v_0=\frac{1}{4}\begin{pmatrix}
0&2&i&3&1&1
\end{pmatrix}^\dagger.
\end{eqnarray}
The corresponding density matrix is $\rho_0=v_0v_0^\dagger$.
 We consider unitary evolution with the Hamiltonian
 \begin{eqnarray}\label{555}
{\mathfrak H}_1=
\begin{pmatrix}
1 &0& i& 2& 0 &1\\
    0 &1& 0& 0& 0& 0 \\
    -i &0& 3& 0& 0 &-4i\\
    2& 0& 0& 4& 0& 0\\
    0 &0 &0 &0 &5& 0\\
    1 &0& 4i& 0& 0& 4\\
    \end{pmatrix}.
\end{eqnarray}
We have explained earlier that our approach to open systems
is an effective formalism and has the parameter $z$ (with $|z|=1$) which is some kind of coupling constant between the system and the external world.

We used Eq.(\ref{98}) to calculate numerically $J(t)$ for  $z=\exp\left (i\frac{\pi}{4}\right )$ in Fig.\ref{f1}, and for $z=\exp\left (i\frac{2\pi}{3}\right )$ in Fig.\ref{f2}.
The results show a quasi-periodic exchange between the system and the environment, with the probability current taking alternately positive and negative values.
Comparison of Fig\ref{f1} with Fig.\ref{f2} shows that $J(t)$ depends on the parameter $z$.

For small $t$ we also used the truncated Taylor series  in Eq.(\ref{100}) (with three terms), to calculate
the probability current $J(t)$. The results are presented in table \ref{t1}, for the Hamiltonian ${\mathfrak H}_1$ and also for a second Hamiltonian
\begin{eqnarray}\label{555A}
{\mathfrak H}_2={\rm diag}
\begin{pmatrix}
1&2&3&4&5&6
\end{pmatrix}
\end{eqnarray}
We also used three values of $z$ ($z=\exp\left (i\frac{\pi}{4}\right )$, $z=\exp\left (i\frac{\pi}{5}\right )$ and $z=\exp\left (i\frac{\pi}{6}\right )$).
The advantage of this approach is that it gives an analytical expression for $J(t)$, which is a quadratic polynomial of $t$.
The disadvantage is that it holds for small $t$.
Of course  if we include many terms (higher order polynomial of $t$), the approximation will improve.

\end{example}

\begin{figure}[!htb]
\centering
\includegraphics[width=8cm]{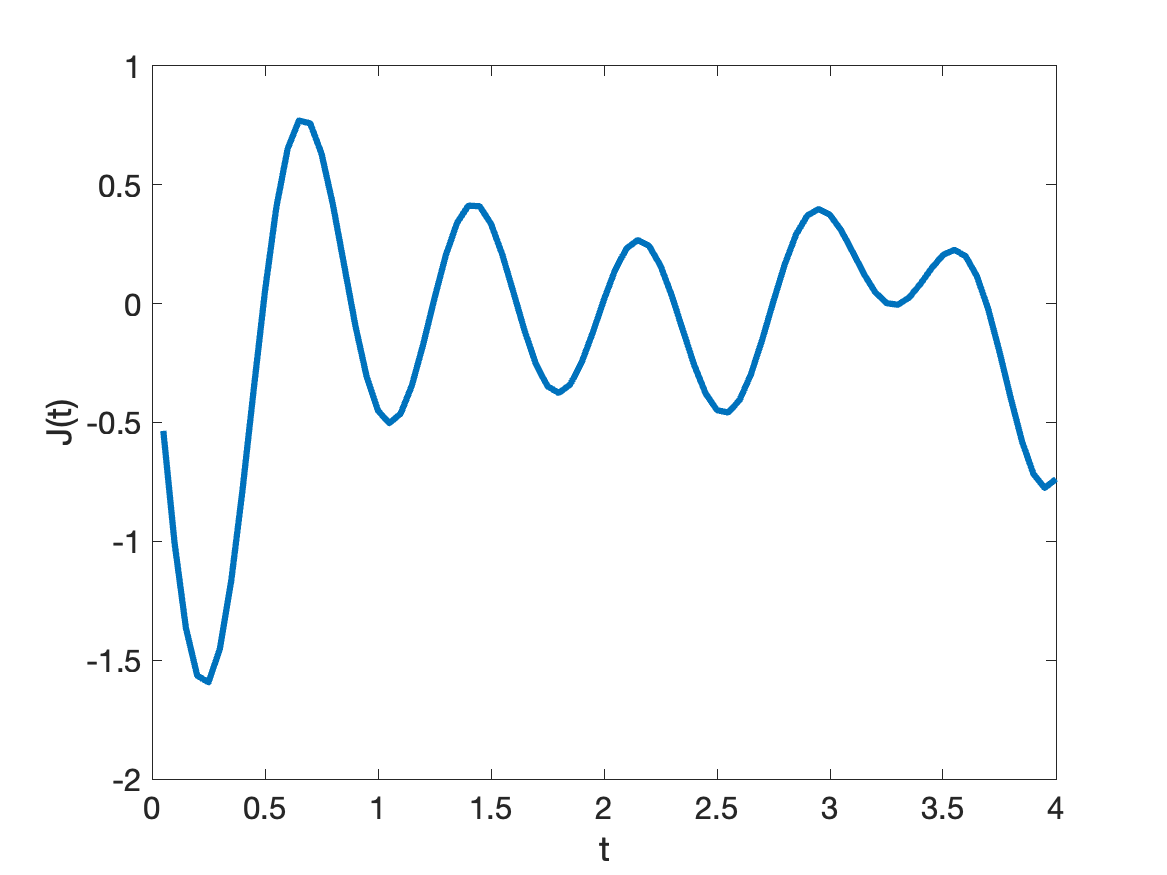}
\caption{The probability current $J(t)$ as a function of the time $t$. At $t=0$ the system is in the mixed state $\frac{1}{3}\Pi(z)$ (Eq.(\ref{P12})), and evolves unitarily with the Hamiltonian ${\mathfrak H}_1$ in Eq.(\ref{555}).The parameter $z=\exp\left (i\frac{\pi}{4}\right )$.}
\label{f3}
\end{figure}

\begin{example}
This is similar to the previous example but with different initial state.
In the `full universe' $\Sigma(6)$, we consider at $t=0$ the density matrix $\rho_0=\frac{1}{3}\Pi(z)$ which is the $z$-Bargmann representation of the density matrix $\frac{1}{3}{\bf 1}_3$ in $\Sigma(3)$
(Eq.(\ref{93A})). 
We used Eq.(\ref{98}) to calculate numerically $J(t)$ for  $z=\exp\left (i\frac{\pi}{4}\right )$.
As in the previous example, the results in Fig.\ref{f3} show a quasi-periodic exchange between the system and the environment.

\end{example}
\subsection{Physical meaning of the probability current $J(t)$ and the parameter $z$}

$J(t)$ is a probability current that flows from the system to the external world. In a specific model (e.g., if we have damping) it could in principle be measured.
$J(t)$ depends on a parameter $z$ that describes the coupling of the system with the external world.

In \cite{VOU4}  there is a specific model  based on quantum tunnelling, where the system is on the left side of the real line, the external world is on the right side, and there is a high potential between them. In that example, the projector $\Pi$ depends on the parameters that describe tunnelling.
Here the formalism is more abstract, but $z$ is still a parameter that describes the coupling of the system to the external world.

We note that:
\begin{itemize}
\item
$z$ is not a gauge-like parameter. It is a physical parameter that describes the coupling of the system with the external world, and $J(t)$ depends on it.
\item
Any model that links two systems has coupling constants (typically many).The equivalent of that here is a single parameter $z$.
\end{itemize}

\section{ Grothendieck ${\cal Q}$ as a figure of merit for open systems}

\subsection{Unitary transformations in $\Sigma(2d)$ are projected as rescaling transformations in $\Sigma(d)$}\label{RRR}

The following proposition shows that embedding of the open system $\Sigma(d)$ into the isolated system $\Sigma(2d)$, introduces in a natural way rescaling matrices into $\Sigma(d)$. 
\begin{proposition}\label{L1}
Let $V$ be a $2d\times 2d$ unitary matrix acting on $\Sigma(2d)$.
Then $\Pi(z)V$ (which acting on a vector $v$ in $\Sigma(2d)$ gives a vector in $\Sigma(d)$) 
and also $\Pi(-z)V$ (which acting on a vector $v$ in $\Sigma(2d)$ gives a vector in $\Sigma_{\rm ext}(d)$),  are both rescaling matrices (they belong in ${\cal S}_{2d}$).
\end{proposition}
\begin{proof}
We use Eq.(\ref{4}) and taking into account the fact that $V$ is a unitary matrix and also that the diagonal elements of $\Pi(z)$ are $\frac{1}{2}$, we get
\begin{eqnarray}
[\Pi(z)VV^\dagger\Pi(z)]_{ii}=[\Pi(z)]_{ii}=\frac{1}{2}.
\end{eqnarray}
This proves that $\Pi(z)V$ are rescaling matrices. The proof for $\Pi(-z)V$ is similar.
\end{proof}

\subsection{The Grothendieck ${\cal Q}$ is linked to probability currents}\label{RR1}

In the isolated system $\Sigma(2d)$ we consider the density matrix $\rho(t)$ in Eq.(\ref{TT}).
Its projection in the open system $\Sigma(d)$ is $\Pi(z) \rho(t)\Pi(z)$, and we will calculate the Grothendieck ${\cal Q}(t)$ in Eq.(\ref{456}) with $\theta=\rho(t)$ and $V=W=\Pi(z)$:
 \begin{eqnarray}
{\cal Q}(t)=\left |{\rm Tr}\left (\frac{\Pi(z)}{{\cal N}[\Pi(z)]}\frac{\rho(t)}{g[\rho(t)]}\frac{\Pi(z)}{{\cal N}[\Pi(z)]}\right )\right |.
\end{eqnarray} 
We have seen in  Eq.(\ref{6BA}) that ${\cal N}[\Pi(z)]=\frac{1}{\sqrt{2}}$ and we get
\begin{eqnarray}\label{890}
{\cal Q}(t)=\frac{2}{g[\rho(t)]} |{\rm Tr}[\Pi(z)\rho(t)] |.
\end{eqnarray} 
We can link the $\frac{d}{dt}{\cal Q}(t)$ with the probability current $J(t)$.
For example if ${\rm Tr}[\Pi(z)\rho(t)]>0$ in some time interval, then Eq.(\ref{700}) gives
\begin{eqnarray}\label{116}
J(t)=\frac{1}{2}\frac{d}{dt}[Q(t)g[\rho(t)]]=\frac{1}{2}\frac{dQ(t)}{dt}g[\rho(t)]+\frac{1}{2}\frac{dg[\rho(t)]}{dt}Q(t).
\end{eqnarray} 
In the special case that $g[\rho(t)]\approx g(\rho_0)$
(i.e., $\frac{d}{dt}g[\rho(t)]\approx 0$) then
\begin{eqnarray}\label{117}
\frac{d}{dt}{\cal Q}(t)\approx \frac{2J(t)}{g(\rho_0)}.
\end{eqnarray} 
Roughly speaking, in open systems the time derivative of ${\cal Q}(t)$ 
is related to the probability current, which means that ${\cal Q}(t)$ is  related to the integral of the probability current.

In the special case that $\Sigma(d)$ is an isolated system, $J=0$ and ${\cal Q}(t)$ does not depend on $t$ (this agree with the third part of proposition \ref{L50}).

\begin{figure}[!htb]
\centering
\includegraphics[width=8cm]{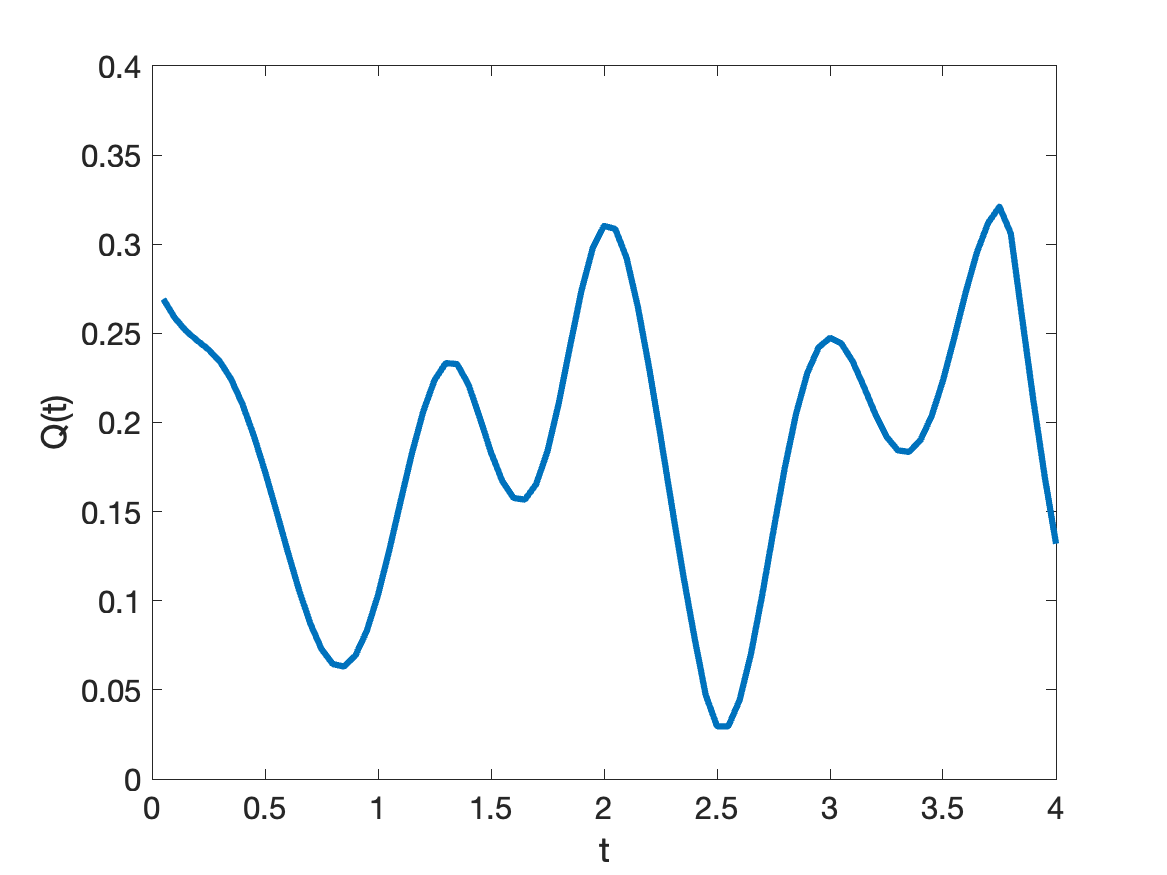}
\caption{${\cal Q}(t)$ as a function of the time $t$. At $t=0$ the the system is in the pure state $v_0$ in Eq.(\ref{55}), and evolves unitarily with the Hamiltonian ${\mathfrak H}_1$ in Eq.(\ref{555}).
The parameter $z=\exp\left (i\frac{\pi}{4}\right )$.}
\label{f4}
\end{figure}

\begin{example}
This example is extension of example \ref{ex10}.
In the isolated system (full universe) $\Sigma(6)$ we consider at $t=0$ the pure state $v_0$ in Eq.(\ref{55}), which evolves unitarily with the Hamiltonian ${\mathfrak H}_1$ in Eq.(\ref{555}).
Then at time $t$, we get $v_t=\exp(i{\mathfrak H}_1t)v_0$ and 
using Eq.(\ref{pure}) (for pure states) we calculate numerically the
\begin{eqnarray}
g[\rho(t)]=g(v_tv_t^\dagger)=||v_tv_t^\dagger||_1.
\end{eqnarray} 
In addition to that we calculate numerically the $|{\rm Tr}[\Pi(z)\rho(t)] |=|v_t^\dagger \Pi(z)v_t|$ for $z=\exp\left (i\frac{\pi}{4}\right )$.
Then we use Eq.(\ref{890}) to calculate ${\cal Q}(t)$ as a function of time.
The results are shown in Fig.\ref{f4}.

We note that the calculation gives
\begin{eqnarray}
Q(0.05)g[\rho(0.05)]=1.1758;\;\;\;Q(0.1)g[\rho(0.1)]=1.2049;\;\;\;Q(0.15)g[\rho(0.15)]=1.2213.
\end{eqnarray} 
\end{example}
Using the central difference method for the derivative, we find
\begin{eqnarray}
\frac{1}{2}\frac{d}{dt}[Q(0.1)g[\rho(0.1)]]=\frac{1}{2}\cdot \frac{1.2213-1.1758}{0.1}=0.227
\end{eqnarray}  
This should be compared with $J(0.1)=0.219$ (from table \ref{t1}). As expected the calculation through the truncated Taylor expansion in Eq.(\ref{100}), gives accurate results for small $t$.

\subsection{Example with ${\cal Q}>1$ in the open system $\Sigma(d)$}\label{RR2}

In the open system $\Sigma(d)$ we consider the density matrix $\frac{1}{d}{\bf 1}_d$ which in the $z$-Bargmann representation is $\frac{1}{d} \Pi(z)$ (Eq.(\ref{93A})).
We show that ${\cal Q}$ in Eq.(\ref{456}) with $\theta=\frac{1}{d} \Pi(z)$ and $V=W=\Pi(z)$, gives values in $(1,k_G)$.
Since ${\cal N}[\Pi(z)]=\frac{1}{\sqrt{2}}$ (Eq.(\ref{6BA})) we get
 \begin{eqnarray}\label{600}
{\cal Q}\left(\frac{1}{d} \Pi(z)\right)=\left | \frac{2{\rm Tr}[\frac{1}{d}\Pi(z)]}{g[\frac{1}{d}\Pi(z)]}\right |=\frac{2d}{g[\Pi(z)]}>1.
\end{eqnarray} 
We used here the strict inequality in Eq.(\ref{400}).

The Grothendieck bound $k_G$ is a very general constraint for Harmonic Analysis and Quantum Mechanics, and examples that go near it are of special interest.
The examples in proposition \ref{L50} for an isolated system (all pure states and all diagonal mixed states), gave ${\cal Q}\le 1$. 
Here we gave an example with ${\cal Q}>1$ in an open system.
Further work is needed to clarify a possible link between open systems and  values of the Grothendieck ${\cal Q}$ in the region  $(1,k_G)$.

\subsection{${\cal Q}$ in open versus isolated systems }\label{sec67}

We consider quantities in $\Sigma(d)$ as an open system within $\Sigma(2d)$, and calculate ${\cal Q}$ for their $z$-Bargmann representatives.
Transformations in $\Sigma(2d)$ are unitary, and their projections in $\Sigma(d)$ are rescaling transformations (proposition \ref{L1}).
We compare this with ${\cal Q}$ for the same quantities, when $\Sigma(d)$ is an isolated system, subject to unitary transformations.
We show that the values of ${\cal Q}$ in these two cases are different.

We start with the following example:
\begin{itemize}

\item
The density matrix $\frac{1}{d}{\bf 1}_d$ in the open system $\Sigma(d)$, whose $z$-Bargmann representative is $\frac{1}{d} \Pi(z)$.
In this case ${\cal Q}>1$ (Eq.(\ref{600})), $\frac{d{\cal Q}}{dt}\ne 0$, and the probability current is $J\ne 0$ (Fig.\ref{f3}).

\item
The density matrix $\frac{1}{d}{\bf 1}_d$ (and other diagonal density matrices) in the isolated system $\Sigma(d)$.
In this case ${\cal Q}<1$ (Eq.(\ref{29})), $\frac{d{\cal Q}}{dt}= 0$ (proposition \ref{L50} third part) and the probability current is $J=0$.

\end{itemize}

We next generalise this.
We replace the various $d\times d$ matrices in the open system $\Sigma(d)$ with their $z$-Bargmann representatives which are $2d\times 2d$ matrices:
\begin{eqnarray}
\theta\rightarrow \theta_1 =[M(z)]^\dagger \theta M(z);\;\;\;
V\rightarrow V_1=[M(z)]^\dagger V M(z);\;\;\;W\rightarrow W_1=[M(z)]^\dagger W M(z)
\end{eqnarray}
Then we get:
 \begin{eqnarray}
{\cal Q}=\left |{\rm Tr}\left (\frac{W^\dagger}{{\cal N}(W)}\frac{\theta}{g(\theta)}\frac{V}{{\cal N}(V)}\right )\right |;\;\;\;
{\cal Q}_1=\left |{\rm Tr}\left (\frac{W_1^\dagger}{{\cal N}(W_1)}\frac{\theta_1}{g(\theta_1)}\frac{V_1}{{\cal N}(V_1)}\right )\right |
\end{eqnarray} 
It is easily seen (using the $M(z)[M(z)]^\dagger={\bf 1}_d$) that
\begin{eqnarray}
{\rm Tr}(W_1^\dagger \theta_1 V_1)={\rm Tr}(W^\dagger \theta V).
\end{eqnarray}
But we have shown in examples (Eqs.(\ref{P0}),(\ref{P1})) that in general $g(\theta_1)\ne g(\theta)$ and ${\cal N}(V_1)\ne {\cal N}(V)$, ${\cal N}(W_1)\ne {\cal N}(W)$. 
Therefore ${\cal Q}\ne {\cal Q}_1$.

\section{Discussion}

Open systems and irreversible phenomena have been studied for a long time using tensor product formalisms.
In this paper we studied an effective direct sum approach to an open system as a subsystem of a larger isolated system, and linked this to probability currents and the Grothendieck formalism.
Our approach involved the following steps.
\begin{itemize}
\item
We expressed the Grothendieck bound formalism in terms of rescaling matrices which change not only the phase but 
also the absolute value of a wavefunction and are suitable for the description of amplification/damping (Eq.(\ref{GR1})).
${\cal Q}$ is a general quantity which in special cases reduces to well known physical quantities, as discussed in section \ref{sec46}.
\item
We introduced semi-unitary matrices in Eq.(\ref{65}), and studied their properties. They are a coherent state formalism in a shorthand language.
\item
We used the semi-unitary matrices to define $z$-Bargman representations (Eq.(\ref{78}) for vectors, and Eq.(\ref{93}) for matrices). 
They embed the Hilbert space $H(d)$ of an open system $\Sigma(d)$, into a larger Hilbert space $H(2d)$ in a natural way.
$H(2d)$ is the space of the full universe which is an isolated system, and is the direct sum of $H(d)$ and $H_{\rm null}(d)$ (Eq.(\ref{780})). $H_{\rm null}(d)$ describes the external world 
to the open system $\Sigma(d)$.
\item
We have introduced probability currents in section \ref{prob} and linked them to the time derivative of the Grothendieck ${\cal Q}$ in Eqs.(\ref{116}), (\ref{117}).
Several examples have shown a quasi-period exchange between the system and the environment, with the probability current taking alternately positive and negative values.
\item
We have shown that unitary transformations in the isolated system $\Sigma(2d)$ reduce to rescaling transformations when projected to its open subsystem $\Sigma(d)$.
\item
We have shown that the values of ${\cal Q}$ and $\frac{d{\cal Q}}{dt}$ for states in an open system, are different from their counterparts in an isolated system (section \ref{sec67}).
\end{itemize}

Immediate applications of the work are in the developing  area of quantum technologies. However, there are interesting applications in other areas also (e.g, in the Physics at Planck scale \cite{NB}). 
Overall, the work is a novel approach to open systems and irreversible phenomena, using direct sum of Hilbert spaces, probability currents and the Grothendieck quantities.\newpage

\begin{table} 
\caption{Summary of the main symbols}
\def\arraystretch{2}
\begin{tabular}{|c|}\hline
${\cal C}(\theta)=\left |\sum_{r,s}\theta _{rs}a_rb_s\right |;\;\;\;|a_r|\le 1;\;\;\;|b_s|\le 1$\\\hline
${\cal Q}(\theta)=\left |\sum_{r,s}\theta_{rs}\lambda_r\mu_s\bra{u_r}v_s\rangle \right |;\;\;\;\lambda_r, \mu_r\le1.$\\\hline
$M(z)=
\begin{pmatrix}
A(z)&A(-z)
\end{pmatrix}$\\\hline
projector $\Pi(z)=[M(z)]^\dagger M(z)$\\\hline
$g(\theta)$: supremum of set of values of ${\cal C}(\theta)$\\\hline
${\cal N}(V)=\max_i\sqrt{\sum_j|V_{ij}|^2}=\max_i\sqrt{(VV^\dagger)_{ii}}$\\\hline
\end{tabular} \label{t0}
\end{table}

\begin{table} 
\caption{ The probability current $J(t)$ for the two Hamiltonians in Eq.(\ref{555}),(\ref{555A}) and for three values of $z$. The initial state at $t=0$ is $v_0$ (given in Eq.(\ref{55})).
The calculation uses the truncated Taylor series in Eq.(\ref{100}) (with three terms) and holds for small $t$.}
\def\arraystretch{2}
\begin{tabular}{|c|c|c|}\hline
&${\mathfrak H}_1$&${\mathfrak H}_2$\\\hline
$z=\exp\left(i\frac{\pi}{4}\right )$&$J(t)=0.508-3.137t+2.552t^2+...$&$J(t)=0.464+0.331t-1.69t^2+...$\\\hline
$z=\exp\left(i\frac{\pi}{5}\right )$&$J(t)=0.401-1.82t+3.573t^2+...$&$J(t)=0.406+0.441t-1.394t^2+...$\\\hline
$z=\exp\left(i\frac{\pi}{6}\right )$&$J(t)=0.325-0.914t+4.206t^2+...$&$J(t)=0.362+0.508t-1.178t^2+...$\\\hline
\end{tabular} \label{t1}
\end{table}

\end{document}